\documentclass[twocolumn,journal]{IEEEtran}

\IEEEoverridecommandlockouts

\usepackage{color}
\usepackage{graphicx}
\usepackage{epstopdf}
\usepackage{amsmath}
\usepackage{amssymb}
\usepackage{algorithm}
\usepackage{algpseudocode}
\usepackage{amsmath}
\usepackage{multirow}
\usepackage{booktabs}
\usepackage{array}
\usepackage{amsthm}
\usepackage{stfloats}
\usepackage{caption}
\usepackage{subfigure}
\usepackage{bm}
\usepackage{setspace}
\usepackage{mdframed}
\usepackage{diagbox}
\usepackage{dashrule}

\newcommand{\be}{\begin{equation}}
\newcommand{\ee}{\end{equation}}
\newcommand{\bea}{\begin{eqnarray}}
\newcommand{\eea}{\end{eqnarray}}
\newcommand{\ba}{\begin{array}}
\newcommand{\ea}{\end{array}}

\renewcommand{\thesubsubsection}{\thesubsection.\arabic{subsubsection}}

\title{Synergizing Wireless Foundation Models and LLMs: A Multi-Task Framework for MA Systems
\thanks{Manuscript received 6 Mar 2026; revised 5 August 2026; accepted 23 October 2026. This work is supported in part by the open research fund of National Mobile Communications Research Laboratory, Southeast University (No. 2025D06), in part by NSFC 62271163, in part by NSFC U25A201852 and NSFC 62231019, in part by NSFC 62371289 and NSFC 62331022. The associate editor coordinating the review of this article and approving it for publication was Dr. J. Zhang. \emph{(Corresponding author: Yang Liu, Wei Wang.)}}
\thanks{H. Zhang, Y. Liu, M. Liu, and W. Wang are with the School of Information and Communication Engineering, Dalian University of Technology, Dalian 116024, China (e-mail: dlutzhc@mail.dlut.edu.cn; yangliu\_613@dlut.edu.cn; liumengzhen@mail.dlut.edu.cn; wangwei2023@dlut.edu.cn).}
\thanks{Q. Wu is with the School of Information Science and Electronic Engineering, Shanghai Jiao Tong University, 200240, China (e-mail: qingqingwu@sjtu.edu.cn)}
\thanks{Q. Shi is with the School of Computer Science and Technology, Tongji University, Shanghai 201804, China, and also with Shenzhen Research Institute of Big Data, Shenzhen 518172, China (e-mail: shiqj@tongji.edu.cn).}
}
\author{\vspace{-20pt} Haocheng Zhang,
Yang Liu, Mengzhen Liu, Qingqing Wu, Qingjiang Shi, and Wei Wang
\vspace{-20pt}
}

\begin{document}
\maketitle
\pagestyle{empty}
\thispagestyle{empty}


\begin{abstract}

While movable antennas (MAs) provide enhanced spatial flexibility through dynamic repositioning, they confront critical challenges in acquiring complete channel state information (CSI) and efficiently optimizing antenna positions.
Encouragingly, wireless foundation models (WFMs) capture generalizable channel priors for high-quality CSI extrapolation, while large language models (LLMs) enable parameter-efficient, token-level optimization for wireless tasks. In this paper, we synergistically integrate a WFM and an LLM into a multi-task framework for MA systems.
First, to address the mismatch between existing WFM pre-training and MA channel characteristics, we develop an MA-oriented ``ExtLWM'' based on the large wireless model (LWM) architecture. Second, we design a unified ExtLWM-Connector-LLM architecture that flexibly adapts to different MA configurations without structural redesign. Third, we reformulate discrete MA positioning as a sequential decision-making problem, where candidate positions are represented by numerical tokens and optimized using group relative policy optimization (GRPO) to exploit the serial generation capability of the LLM. Finally, a multi-stage training strategy is developed to maintain multi-task performance under limited training samples while mitigating catastrophic forgetting.
Simulation results show that ExtLWM can produce high-quality channel embeddings while outperforming various benchmarks in CSI extrapolation. Meanwhile, the proposed LLM-based approach achieves a favorable trade-off between performance and computational complexity.

\end{abstract}


\section{Introduction}\label{section:1}

Movable antenna (MA) is an emerging technology for next-generation wireless communications, providing a new approach to further enhance wireless transmission capability \cite{CM Zhu}. Specifically, unlike conventional fixed-position antenna (FPA) arrays, whose elements are fabricated at predetermined locations, MAs are connected to radio-frequency chains through flexible cabling and can be programmably moved via precise motor control \cite{TWC Zhu}. This position reconfigurability introduces additional spatial degrees of freedom (DoFs), allowing MA systems to exploit geometry gain and spatial diversity to improve communication performance without the need for substantial array-size expansion \cite{TWC Zhu}, \cite{TWC Ma}.

Existing MA systems typically adopt either continuous or discrete positioning mechanisms. Although continuous-position adjustment provides greater spatial flexibility and potentially higher performance, it poses significant implementation challenges in practice~\cite{WC Ning}. First, the achievable positioning step size is fundamentally restricted by the physical resolution of mechanical actuators, necessitating sophisticated, high-precision hardware to approximate continuous movement~\cite{TWC Dong}, \cite{JSTSP Shao}. Second, continuous position adjustment typically entails frequent repositioning, resulting in elevated actuator power consumption, prolonged movement latency, and substantial control overhead~\cite{TCOM Wu}. Consequently, discrete-position adjustment offers a more favorable trade-off between system performance and implementation complexity.

Despite its practical advantages, discrete-position MA system will inevitably confront two issues in deployment: The difficulty of acquiring complete CSI from limited channel observations and the complexity of discrete MA position optimization under implementation constraints.
Firstly, for CSI acquisition, due to the limited movement speed of MAs and the constrained pilot overhead, only a small subset of candidate positions can be explored within a given channel coherence interval, resulting in incomplete CSI observations~\cite{WC Ning}. To address this issue, a feasible approach is to extrapolate the CSI of all unobserved candidate positions from incomplete CSI. In \cite{ICC Li}, the authors design an extrapolation algorithm by learning a low-dimensional channel basis and estimating the corresponding latent coefficients. Nevertheless, such model-based methods heavily rely on predefined channel priors. In contrast, data-driven deep neural network (DNN)-based methods can directly learn spatial channel correlations among different candidate positions from large-scale CSI samples, thereby enabling more flexible CSI extrapolation. For instance, the authors of \cite{WCL Zhang} design an asymmetric architecture consisting of a self-attention-based encoder and a graph-attention-based decoder. In \cite{JSAC Gao}, the authors augment the encoder with a mixture-of-experts module to enhance feature extraction and noise resilience. Besides, similar studies~\cite{ICC Wu},~\cite{TWC Wang} have demonstrated that masked autoencoder-based architectures can achieve superior CSI extrapolation performance compared with conventional network architectures. Although the above dedicated CSI extrapolation models can recover complete CSI to directly support downstream tasks~\cite{TWC Wang},~\cite{ICCC Lu}, their training typically relies on sufficient complete CSI samples and their transferability to unseen scenarios remains limited.

Wireless foundation models (WFMs), which learn generalizable channel representations through large-scale pre-training, offer a promising avenue to mitigate these challenges by delivering superior sample efficiency across diverse downstream tasks \cite{Arxiv Alikhani}--\cite{PIMRC Ghassemi}.
The large wireless model (LWM) proposed in \cite{Arxiv Alikhani} achieves near-saturated line-of-sight (LoS)/non-LoS (NLoS) identification performance with less than 1\% labeled samples, while attaining comparable beam prediction performance using only 40\%-50\% of the training data.
Regarding cross-scenario transferability, the ContraWiMAE proposed in \cite{JSAC Guler} maintains competitive performance when transferred to unseen propagation scenarios and carrier frequencies, further demonstrating the generalizability of its pre-trained channel representations.
These advantages are particularly valuable for MA systems, where acquiring sufficient complete CSI samples across candidate positions is often impractical, while channel characteristics can vary substantially across different propagation environments and position configurations. Nevertheless, the use of WFMs for CSI extrapolation in MA systems, particularly under highly incomplete observations and diverse scenarios, remains insufficiently explored.

Secondly, discrete MA position optimization requires selecting the location of each MA from a large number of candidate points while satisfying inter-MA distance constraints, giving rise to a computationally challenging combinatorial optimization problem. Conventional branch-and-bound (BnB) methods can obtain high-quality solutions but suffer from prohibitive computational complexity as the search space grows~\cite{TCOM Wu}, whereas penalty-based relaxation (PBR) methods reduce the optimization burden at the cost of non-negligible performance loss~\cite{CL Cang}. Besides, network-based approaches also face inherent limitations in handling such discrete combinatorial structures, since directly formulating the problem as a classification task leads to an excessively large output space, whereas generating continuous positions followed by discretization may result in suboptimal or even infeasible solutions~\cite{TNSE Guo},~\cite{TWC Zhang}. Overall, existing approaches still suffer from certain limitations in terms of computational efficiency and solution quality.

To strike a more favorable balance, the combinatorial MA position optimization problem can be reformulated as a sequential optimization problem. By decomposing the high-dimensional joint decision into a sequence of lower-dimensional decisions, sequential optimization can significantly alleviate the curse of dimensionality and substantially reduce optimization complexity~\cite{TWC Zhang}. Meanwhile, compared with parallel feedforward computation, sequential optimization can provide greater effective computational depth, and thus has the potential to achieve better performance on problems with strong sequential dependencies under comparable model sizes and training data~\cite{ICLR Liu},~\cite{CoRL Chebotar}.
Notably, large language model (LLM) inherently follows an autoregressive generation mechanism, which is well aligned with the sequential decision structure of the reformulated MA positioning problem. In addition to this structural compatibility, LLMs can be adapted to wireless communication tasks through lightweight task-specific training with only a limited number of trainable parameters. Existing studies have demonstrated such parameter-efficient adaptation~\cite{TCOM Xu}-\cite{TCOM Zhang}. For example, the authors of \cite{TCOM Xu} fine-tune only about 5.95M parameters of GPT-2 and achieve sum-rate performance comparable to that of a 13.15M-parameter transformer trained from scratch. Moreover, the authors of \cite{WCL Sheng} propose a prompt-as-prefix strategy that converts historical optimal beam indices and angle information into discrete numerical tokens, thereby improving the average normalized beamforming gain of GPT-2 by approximately 62.4\% for the beam prediction task.
However, existing studies mainly focus on parameter-efficient adaptation of pre-trained LLMs to individual wireless tasks, while their intrinsic autoregressive capability for sequential optimization in wireless communications remains largely unexplored. This gap motivates the use of pre-trained LLMs as sequential decision models for discrete MA position optimization.

More importantly, CSI extrapolation and MA position optimization should not be treated as isolated tasks, but rather jointly constitute an integrated wireless communication process. Developing separate task-specific models would lead to duplicated representation learning, inefficient use of limited training samples, and limited scalability across different MA configurations. To enable unified and efficient operation throughout this entire process, synergizing WFMs and LLMs offers a promising multi-task paradigm. Specifically, WFMs can perform reliable CSI extrapolation and extract informative feature embeddings from limited training samples, while LLMs can leverage these representations to support both classification and sequential optimization. However, effectively integrating these two types of foundation models into a unified architecture remains nontrivial, and systematic studies on multi-task MA systems are still limited.

Motivated by the above analysis, this paper innovatively integrates the WFM and LLM to support multiple downstream tasks in MA systems. Specifically, we consider three representative tasks, including CSI extrapolation as a regression task, LoS/NLoS identification as a classification task, and sum-rate maximization as a sequential optimization task. Building upon this framework, this paper aims to explore the under-investigated network architectures and training paradigms for synergizing WFMs and LLMs in MA systems, and seeks to answer the following pivotal questions: Q1)~Can existing WFMs developed for complete CSI in FPA systems and primarily designed for feature extraction be effectively transferred to MA systems? In other words, is it necessary to develop a dedicated WFM for MA systems under incomplete CSI? Q2)~How can WFMs and LLMs be seamlessly integrated at the network architecture level to construct a unified framework with a fixed backbone? Q3)~How can the potential of LLMs for sequential optimization of discrete MA positions be effectively exploited? Q4)~How can a unified training paradigm be designed under limited training samples while maintaining satisfactory performance across different tasks? Correspondingly, the main contributions of this paper can be summarized as follows:

\begin{itemize}
\item For Q1), we develop a WFM for MA systems under incomplete CSI, termed ``ExtLWM''. ExtLWM can simultaneously provide high-quality extrapolated CSI and informative feature embeddings. Simulation results demonstrate that ExtLWM exhibits strong generalization across different scenarios, CSI incompleteness ratios, orthogonal frequency division multiplexing (OFDM) configurations, and array structures. Moreover, its CSI extrapolation performance outperforms the FPA-oriented LWM-1.1 and models trained from scratch, with more pronounced advantages under limited training data.

\item For Q2), we design a lightweight Q-Former-based connector to seamlessly integrate the WFM and LLM. Meanwhile, representative tasks in MA systems are unified within the proposed framework: CSI extrapolation is handled by the WFM, while LoS/NLoS identification and sum-rate maximization are further formulated as token-level optimization tasks for the LLM. Simulation results demonstrate that the proposed framework can effectively accommodate different MA system configurations, and also exhibits good compatibility with different LLMs.

\item For Q3), we formulate the sequential MA positioning problem within a Markov decision process (MDP) framework and represent MA positions using numerical tokens. Building on this formulation, we employ a group relative policy optimization (GRPO) algorithm for optimization. Simulation results demonstrate that the proposed method can effectively exploit the capability of LLMs while alleviating the credit-assignment challenge. Moreover, it achieves higher sum-rate performance than conventional PBR methods with low execution overhead.

\item For Q4), we propose a multi-stage training algorithm for data-limited scenarios. In particular, to preserve the performance of the classification task while optimizing the sum-rate, we incorporate experience replay (ER) into GRPO training process. Simulation results demonstrate that the proposed multi-stage training strategy achieves better performance across multiple tasks than joint training, while the incorporation of ER more effectively mitigates catastrophic forgetting.

\end{itemize}

\vspace{-0.2cm}

\section{System Model and Problem Formulation} \label{section:2}

\begin{figure*}[t]
\centering
\includegraphics[width = 7 in]{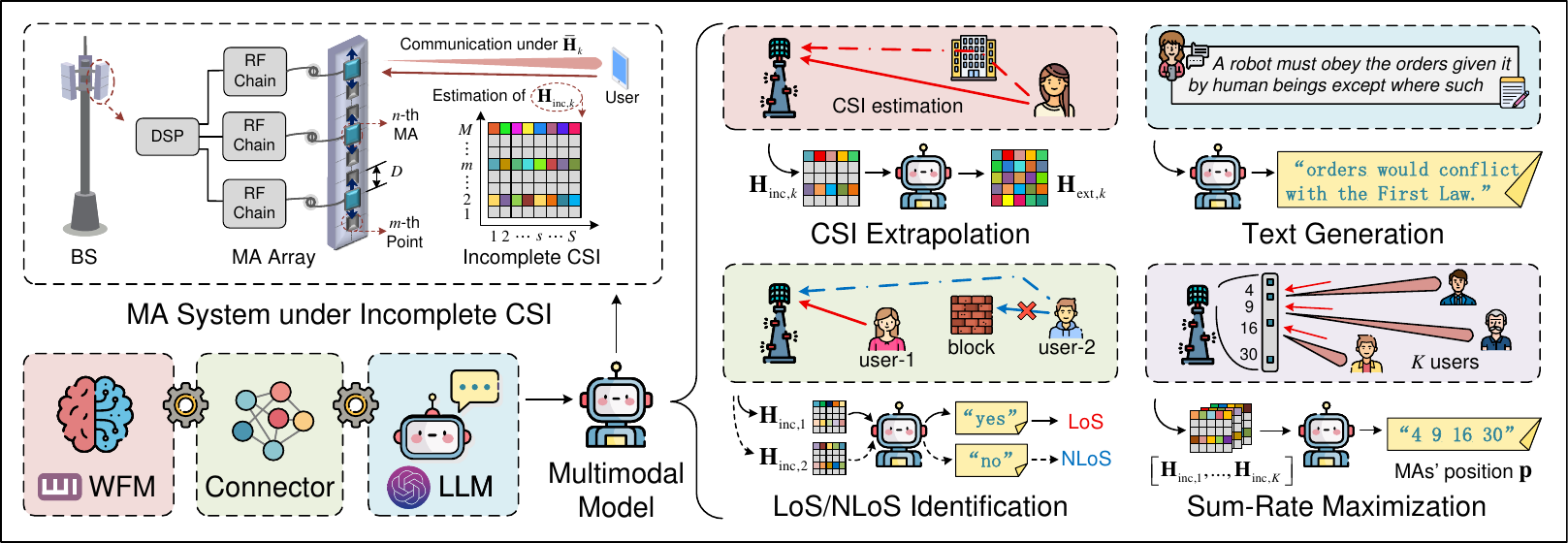}
\caption{Overview of the MA system under incomplete CSI, where a multimodal model simultaneously enables multiple tasks.}
\label{fig_system}
\vspace{-0.3 cm}
\end{figure*}

We consider an MA-enabled OFDM downlink communication system, where a BS is equipped with $N$ MAs to serve $K$ single-antenna users. As shown in Fig.~\ref{fig_system}, the BS is equipped with a one-dimensional array featuring $M$ candidate points spaced by distance $D$ that are available for the deployment of $N$ antenna elements ($M \gg N$). Let $\mathcal{M}\triangleq\{1,2,...,M\}$, $\mathcal{N}\triangleq\{1,2,...,N\}$ and $\mathcal{K}\triangleq\{1,2,...,K\}$ denote the sets of all candidate positions, MAs and users, respectively. Define position vector $\mathbf{p}\triangleq[p_1,p_2,...,p_N]^T$ to indicate the position indices of $N$ MAs with the entries $p_n\in\mathcal{M}$ arranged in ascending order, i.e., $p_i>p_j$ for $i>j$. Besides, to mitigate mutual coupling effect, a minimum separation distance should be maintained between MAs, i.e.,
\begin{equation}
\label{eq_distance}
 p_i - p_j  \geq \frac{d_\mathrm{min}}{D}, \forall i, j \in \mathcal{N}, i > j,
\end{equation}
where $d_\mathrm{min}$ is the ``safety spacing'' with $d_\mathrm{min} \geq D$. The BS employs OFDM modulation, where $\mathcal{S}\triangleq\{1,2,...,S\}$, $\lambda_{\mathrm{c}}$, $f_{\mathrm{c}}=\frac{c_0}{\lambda_{\mathrm{c}}}$, and $\Delta f$ represent the subcarrier set, center carrier wavelength, center carrier frequency, and subcarrier spacing, respectively. The frequency of the $s$-th subcarrier is $f_s =f_{\mathrm{c}}+(s-\frac{S+1}{2}) \Delta f$.

\vspace{-0.2cm}

\subsection{Channel Model and Incomplete Observation}
The downlink channel is described by a multi-carrier geometry-based channel model to characterize multipath propagation, which is consistent with the modeling principles adopted by state-of-the-art ray-tracing-based channel simulators/datasets, such as DeepMIMO~\cite{ITA Alkhateeb} and Sionna~\cite{Arxiv Aoudia}. Specifically, define $\mathbf{h}_{k,s} \in \mathbb{C}^{M}$ as the channel vector between the $k$-th user and all $M$ candidate locations at subcarrier $s$, which is modeled as
\begin{equation}
\label{eq_complete_csi}
\mathbf{h}_{k,s} =\sum_{g=1}^{G_k} \sqrt{\frac{1}{\rho_{k, g}}} \alpha_{k, g} e^{-j 2 \pi f_s \tau_{k, g}} \mathbf{a}(\varphi_{k, g}),
\end{equation}
where $G_k$ denotes the number of effective propagation paths for user $k$, while $\rho_{k,g}$, $\alpha_{k,g}$, $\tau_{k,g}$, and $\varphi_{k,g}$ denote the path loss, frequency-independent complex path coefficient, propagation delay, and angle of departure (AoD) of the $g$-th path of user $k$, respectively. Moreover, $\mathbf{a}(\varphi_{k,g})$ denotes the array steering vector corresponding to the $M$ candidate MA positions, which is further expressed as
\begin{equation}
\label{eq_complete_vector}
\mathbf{a}(\varphi) = [1, e^{-\frac{j 2 \pi f_{\mathrm{c}} D \sin \varphi}{c_0}}, ..., e^{ -\frac{ j 2 \pi f_{\mathrm{c}} (M-1)D \sin \varphi}{c_0}}]^T.
\end{equation}
where $c_0$ denotes the speed of light. Therefore, the complete channel matrix of the $k$-th user is denoted as $\mathbf{H}_k \triangleq [\mathbf{h}_{k,1}, \mathbf{h}_{k,2}, ..., \mathbf{h}_{k,S} ] \in \mathbb{C}^{M \times S}$. Subsequently, denote $[\mathbf{h}_{k,s}]_m$ as the channel coefficient associated with the $m$-th candidate position, i.e., the $m$-th element of $ \mathbf{h}_{k, s} $. Then, the downlink channel vector between $N$ MAs with positions $\mathbf{p}$ and the $k$-th user at the $s$-th subcarrier can be expressed as
\begin{equation}
\label{eq_ma_vector}
\bar{\mathbf{h}}_{k, s} \triangleq \big[\mathbf{h}_{k,s}\big]_{\{\mathbf{p}\}} \in \mathbb{C}^{N}.
\end{equation}

Consequently, the channel matrix spanning $N$ MAs across all subcarriers is denoted by $\bar{\mathbf{H}}_k\triangleq [\mathbf{\bar{h}}_{k,1}, \mathbf{\bar{h}}_{k,2}, ..., \mathbf{\bar{h}}_{k,S} ] $, which is indeed a submatrix of the complete CSI $\mathbf{H}_k$. Conventional studies \cite{TCOM Wu}, \cite{TNSE Guo} typically assume perfect knowledge of complete CSI across all candidate positions, often acquired by concatenating channel measurements across multiple time slots. However, obtaining comprehensive CSI remains practically daunting due to prohibitive pilot overhead and stringent demands on antenna mechanics for instantaneous physical relocation of antenna elements. To be practical, we define the more accessible incomplete CSI for the $k$-th user as $\mathbf{H}_{\mathrm{inc},k} \triangleq [\mathbf{h}_{\mathrm{inc},k,1}, \mathbf{h}_{\mathrm{inc},k,2}, \dots, \mathbf{h}_{\mathrm{inc},k,S}] \in \mathbb{C}^{M \times S}$, where each subcarrier vector is formulated as
\begin{equation}
\label{eq_incomplete_csi}
\mathbf{h}_{\mathrm{inc}, k, s} = \mathbf{h}_{k, s} \odot \mathbf{o},
\end{equation}
where $\mathbf{o} \triangleq [o_1, o_2, \dots, o_M]^T \in \mathbb{R}^{M}$ represents an indicator vector with $o_m \in \{0, 1\}$ indicating whether the channel coefficient of position $m$ is observable. Furthermore, we define the CSI incompleteness ratio as
$\eta_{\mathrm{msk}}\triangleq \frac{1}{M}\sum_{m=1}^{M}(1-o_m)$,
which represents the proportion of candidate positions with unobservable CSI.

\vspace{-0.2cm}
\subsection{CSI Extrapolation}

Given that complete CSI is indispensable for MA configuration, it is highly desirable to reconstruct complete CSI matrix. Specifically, we aim to extrapolate the CSI of unobserved antenna positions based on the sparse measurements through inference, instead of conventional physical scan \cite{JSAC Gao}--\cite{ICC Wu}. Extrapolation of the channel coefficients of unseen positions (i.e., those with $o_j = 0$) reads
\begin{equation}
\label{eq_reconstruct_csi}
\mathbf{h}_{\mathrm{ext}, k, s} =  \bm{f}_{\mathrm{ext}} (\mathbf{h}_{\mathrm{inc},k, s}),
\end{equation}
where $\bm{f}_{\mathrm{ext}}(\cdot)$ is the extrapolation function and $\mathbf{h}_{\mathrm{ext}, k, s}$ is a prediction of the true channel $\mathbf{h}_{k,s}$. Accordingly, let $\mathbf{H}_{\mathrm{ext}, k}\triangleq [\mathbf{h}_{\mathrm{ext}, k, 1}, \mathbf{h}_{\mathrm{ext}, k, 2},..., \mathbf{h}_{\mathrm{ext}, k, S} ]$ represent the extrapolated CSI matrix for user $k$, whose accuracy can be evaluated by normalized mean square error (NMSE) defined as

\begin{equation}
\label{eq_problem_csi}
\mathrm{NMSE}_{k} =  \mathbb{E}_{s \in \mathcal{S}} \bigg[ \frac{\|\mathbf{h}_{\mathrm{ext}, k, s}-\mathbf{h}_{k, s} \|^2}{ \|\mathbf{h}_{k, s} \|^2} \bigg].
\end{equation}

\vspace{-0.2cm}

\subsection{LoS/NLoS Identification}
LoS/NLoS identification is an important component of wireless environment awareness and facilitates context-aware tasks such as parameter estimation and positioning~\cite{ICCC Lu}. Leveraging the acquired incomplete CSI, the identification process for the $k$-th user is formulated as
\begin{equation}
\label{eq_los_cls}
\widehat{\delta}_k =  \bm{f}_{\mathrm{id}} (\mathbf{H}_{\mathrm{inc}, k}),
\end{equation}
where $\bm{f}_{\mathrm{id}}(\cdot)$ represents the identification algorithm/network, while $\widehat{\delta}_k$ and $\delta_k \in \{0, 1\}$ signify the classified and genuine CSI category for user $k$, with $1$ and $0$ indicating LoS and NLoS conditions, respectively. To assess the identification capability across $K$ users, we adopt the weighted F1-score, denoted as $\mathrm{F1}\text{-}\mathrm{score}(\mathcal{K}) \in [0, 1]$, as the performance metric, which jointly accounts for class-wise precision and recall while weighting each class according to its sample proportion, making it suitable for evaluating performance under imbalanced channel-type distributions.

\vspace{-0.2cm}

\subsection{Sum-Rate Maximization}

The sum-rate across all users and subcarriers is a key metric for MA-enabled multi-user communication systems. Let $\mathbf{w}_{k, s} \in \mathbb{C}^{N} $  denote the precoder of the $k$-th user on subcarrier $s$. The received signal by the $k$-th user on subcarrier $s$ can be written as
\begin{equation}
\label{eq_receive_signal}
y_{k,s}=\mathbf{\overline{h}}_{k, s}^H \mathbf{W}_{s} \mathbf{x}_s +n_s,
\end{equation}
where $\mathbf{W}_s \triangleq [\mathbf{w}_{1,s}, \mathbf{w}_{2,s}, ..., \mathbf{w}_{K,s} ] \in \mathbb{C}^{N \times K}$ denotes the precoding matrix for subcarrier $s$, vector $\mathbf{x}_s$ denotes uncorrelated information symbols following complex Gaussian distribution with zero-mean and unit variance, i.e., $\mathbb{E}\{\mathbf{x}_s\mathbf{x}_s^H\}=\mathbf{I}_K$, and $n_s \sim\mathcal{C}\mathcal{N}(0, \sigma_{\mathrm{c}}^2)$ represents the receiving noise. Consequently, the signal-to-interference-plus-noise ratio (SINR) of the $k$-th user on subcarrier $s$ is given as
\begin{equation}
\label{eq_SINR}
\mathrm{SINR}_{k, s} =\frac{|\mathbf{\overline{h}}_{k,s}^H \mathbf{w}_{k, s} |^2}{\sum_{j \neq k} |\mathbf{\overline{h}}_{k, s}^H \mathbf{w}_{j, s} |^2+\sigma_{\mathrm{c}}^2}.
\end{equation}

Subsequently, the achievable communication rate of the $k$-th user on subcarrier $s$ is given by
\begin{equation}
\label{eq_rate}
R_{k,s} =\log _2 (1+\mathrm{SINR}_{k,s} ).
\end{equation}

Recalling \eqref{eq_ma_vector}, the sum-rate of all $K$ users at all $S$ subcarriers, i.e.,$\sum_{s=1}^{S}  \sum_{k=1}^{K} R_{k, s}$, is jointly determined by the positions of MAs and beamforming. Therefore, the sum-rate maximization task under incomplete CSI can be expressed as $ \{ \mathbf{p},  \mathbf{W}_1, ..., \mathbf{W}_S \} =  \bm{f}_{\mathrm{max}} (\{\mathbf{H}_{\mathrm{inc}, 1}, ..., \mathbf{H}_{\mathrm{inc}, K}\})$.

\vspace{-0.2cm}
\subsection{Problem Formulation}

In this paper, instead of relying on traditional task-specific architectures that design $\bm{f}_{\mathrm{ext}}(\cdot)$, $\bm{f}_{\mathrm{id}}(\cdot)$, and $\bm{f}_{\mathrm{max}}(\cdot)$ separately, we propose a unified multimodal network with parameter $\bm{\theta}$ to perform all tasks concurrently. Therefore, the multi-task optimization problem, centered on sum-rate maximization, is formulated as follows
\begin{subequations}\label{eq_problem}
\begin{align}
\max _{\bm{\theta}}  & \, \,  \sum_{s=1}^{S}  \sum_{k=1}^{K} R_{k, s}, \label{eq_max_rate}  \\
\mathrm{ s.t. } & \,\,\,    p_i - p_j  \geq \frac{d_\mathrm{min}}{D}, ~\forall i, j \in \mathcal{N}, i > j,  \label{eq_constraint_pos} \\
& \,\,\, p_n \in \mathcal{M}, ~\forall n \in\mathcal{N},  \label{eq_constraint_p} \\
& \,\,\, \mathrm{NMSE}_{k} \leq \epsilon_{\mathrm{NMSE}}, ~\forall k \in\mathcal{K}, \label{eq_constraint_nmse} \\
& \,\,\, \mathrm{F1}\text{-}\mathrm{score}(\mathcal{K}) \geq \epsilon_{\mathrm{F1}},  \label{eq_constraint_f1} \\
& \, \sum_{s=1}^{S}\sum_{k=1}^{K}
\|\mathbf{w}_{k,s}\|^2
\leq P_{\mathrm{tx}}, \label{eq_constraint_power}
\end{align}
\end{subequations}
where \eqref{eq_constraint_pos} and \eqref{eq_constraint_p} specify the constraints on the MA positions, \eqref{eq_constraint_nmse} and \eqref{eq_constraint_f1} mandate performance thresholds for CSI extrapolation and LoS/NLoS identification, respectively, and \eqref{eq_constraint_power} imposes the power constraint with fixed transmit power $P_{\mathrm{tx}}$.

In this paper, we synergistically integrate the WFM and LLM to construct a unified network parameterized by $\bm{\theta}$, which flexibly accommodates varying numbers of MAs and candidate positions. Specifically, the WFM supports the effective use of the LLM in sample-limited environments by providing high-quality channel embeddings and accurate CSI extrapolation. The LLM then leverages the wireless domain knowledge provided by the WFM to formulate different wireless tasks as token-level optimization problems. Besides supporting conventional classification tasks, the LLM can fully exploit its sequential generation capability for the MA positioning.

\vspace{-0.3cm}

\section{Unified network architecture}\label{section:3}

In this section, we introduce the unified framework that synergistically integrates the WFM and LLM via a connector, i.e., $\bm{\theta} \triangleq \{ \bm{\theta}_{\mathrm{WFM}}, \bm{\theta}_{\mathrm{Con}}, \bm{\theta}_{\mathrm{LLM}}\}$. The detailed architectures of the individual network components are introduced in the following subsections.

\vspace{-0.2cm}

\subsection{ExtLWM-based WFM}\label{section:3-A}

\begin{figure*}[t]
\centering
\includegraphics[width = 7in]{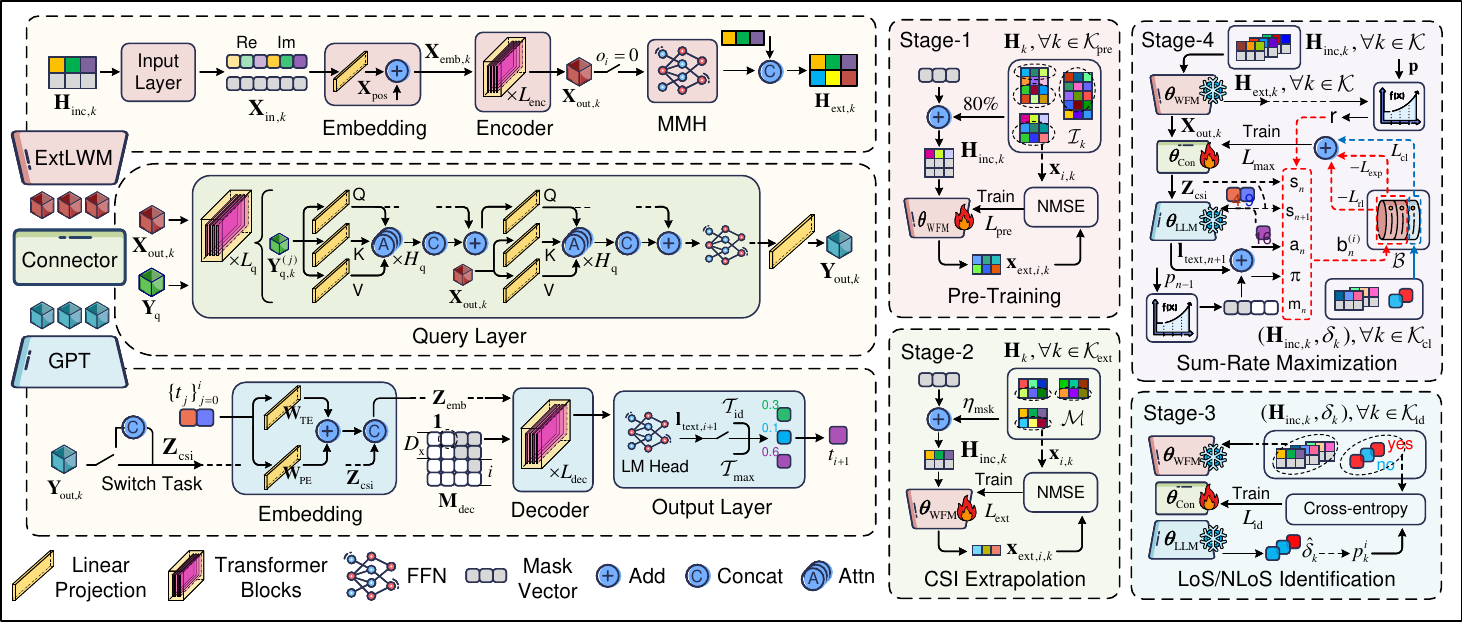}
\caption{Proposed unified network architecture, with the multi-stage training strategy for supporting multiple tasks.}
\label{fig_network}
\vspace{-0.3 cm}
\end{figure*}

LWM has been demonstrated to be effective in extracting channel features for OFDM systems and capable of facilitating diverse downstream physical-layer applications. Considering the mismatch between existing WFMs and MA systems, we develop an MA-oriented WFM based on the LWM architecture, termed ``Extended LWM (ExtLWM)''. ExtLWM enables high-quality CSI extrapolation from incomplete CSI while simultaneously producing informative channel embeddings to support subsequent downstream tasks. Defining $\bm{\theta}_{\mathrm{WFM}}$ as the parameters of ExtLWM, the designs of its components (as illustrated in Fig.~\ref{fig_network}) are specified in the following.

\subsubsection{Input Layer}
The input to ExtLWM consists of incomplete CSI measurements $\mathbf{H}_{\mathrm{inc},k}$ collected at candidate positions whose locations are known. These measurements are organized into a formatted data matrix $\mathbf{X}_{\mathrm{in},k} \in \mathbb{R}^{M\times 2S}$, which comprises $M$ position-wise patches.  Specifically, for $n\in\mathcal{N}$, the $p_n$-th patch $\mathbf{x}_{p_n,k}^T$ of $\mathbf{X}_{\mathrm{in},k}$ is given as
\begin{equation}
\begin{aligned}\label{eq_patch}
\mathbf{x}_{p_n,k}^T=\alpha_\mathrm{p} \big[&\Re\{h_{\mathrm{inc},p_n,k,1}\},...,\Re\{h_{\mathrm{inc},p_n,k,S}\}, \\
&\Im\{h_{\mathrm{inc},p_n,k,1}\},...,\Im\{h_{\mathrm{inc},p_n,k,S}\} \big],
\end{aligned}
\end{equation}
where $h_{\mathrm{inc}, p_n,k,s}$ is the channel coefficient of the $n$-th MA over the $s$-th subcarrier, i.e., $\{\mathbf{h}_{\mathrm{inc},k,s}\}_{p_n}$,  and $\alpha_\mathrm{p}$ is a scaling parameter. Moreover, the CSI patches corresponding to vacant candidate positions are replaced with a predefined $[\mathrm{MASK}]$ value\footnote{$[\mathrm{MASK}]$ serves as a unified representation of missing information in ExtLWM. Specifically, it is used to randomly mask CSI from multiple scenarios during pre-training, to represent incomplete CSI in downstream tasks, and to pad missing subcarriers. Therefore, ExtLWM can naturally accommodate channel inputs with fewer subcarriers than the pre-training configuration $S$.}, i.e.,
\begin{equation}
\label{eq_mask}
\mathbf{x}_{i,k}=[\mathrm{MASK}]\cdot \mathbf{1}_{2S}^T,  \; \mathrm{s.t.} \; [\mathbf{o}]_i=0, \forall i \in \mathcal{M},
\end{equation}
where $\mathbf{o}$ is the indication vector (recall \eqref{eq_incomplete_csi}).

\subsubsection{Embedding Layer}
The embedding layer projects the masked CSI data into a high-dimensional space, facilitating the encoder in discovering hidden features related to spatial correlation among antenna positions as well as multipath and scattering characteristics. The signal flow of the embedding layer can be expressed as follows
\begin{equation}
\label{eq_lwm_emb}
\mathbf{X}_{\mathrm{emb},k} = \operatorname{LN}\big(\operatorname{Linear}(\mathbf{X}_{\mathrm{in},k}) + \mathbf{X}_\mathrm{pos}\big),
\end{equation}
where the linear mapping $\operatorname{Linear}(\cdot)$ elevates the signal dimension from $M\times 2S$ to $M\times D_{\mathrm{LWM}}$ and $\operatorname{LN}(\cdot)$ denotes the layer normalization operation, $\mathbf{X}_{\mathrm{pos}}\in\mathbb{R}^{M\times D_{\mathrm{LWM}}}$ denotes the position embedding of $M$ patches that encodes position information into signal.

\subsubsection{Encoder Layer}

The encoder layer aims to extract high-dimensional features from incomplete CSI embeddings, thereby establishing a robust representation capable of capturing intricate intrinsic correlations. Inspired by the architecture of LWM-1.1 \cite{Arxiv Alikhani}, the encoder layer consists of $L_{\mathrm{enc}}$ cascaded encoder blocks with identical structures. The $j$-th encoder block takes $\mathbf{X}_{\mathrm{enc},k}^{(j)}$ as the input and outputs the enhanced representation $\mathbf{X}_{\mathrm{enc},k}^{(j+1)}$, where the initial input is given by $\mathbf{X}_{\mathrm{enc},k}^{(1)}=\mathbf{X}_{\mathrm{emb},k}$.

Each block integrates a multi-head attention module followed by a feed-forward network (FFN) module. Specifically, within the $j$-th block, the input $\mathbf{X}_{\mathrm{enc},k}^{(j)}$ is first projected onto three independent linear subspaces to generate the query, key, and value matrices, denoted as $\mathbf{Q}_{\mathrm{enc},k}^{(j)}$, $\mathbf{K}_{\mathrm{enc},k}^{(j)}$, and $\mathbf{V}_{\mathrm{enc},k}^{(j)} \in \mathbb{R}^{M \times D_{\mathrm{LWM}}}$, respectively. To enable parallel processing of distinct feature subspaces, each matrix is evenly partitioned into $H_{\mathrm{enc}}$ heads. The scaled dot-product attention for the $h$-th head is computed as
\begin{equation}
\label{eq_mha_lwm}
\hspace{-0.03cm} \mathbf{X}_{\mathrm{enc}, k, h}^{(j)} = \operatorname{Softmax}\Bigg(\frac{\mathbf{Q}_{\mathrm{enc}, k, h}^{(j)}\big( \mathbf{K}_{\mathrm{enc}, k, h}^{(j)} \big)^T}{\sqrt{ \frac{D_{\mathrm{LWM}}}{H_{\mathrm{enc}}}}}\Bigg)\mathbf{V}_{\mathrm{enc}, k, h}^{(j)},
\end{equation}
where $\mathbf{X}_{\mathrm{enc}, k, h}^{(j)} \in \mathbb{R}^{M \times \frac{D_{\mathrm{LWM}}}{H_{\mathrm{enc}}}}$. Subsequently, the outputs of all attention heads are concatenated, linearly transformed, and further processed by the FFN, yielding the output of the $j$-th encoder block, $\mathbf{X}_{\mathrm{enc},k}^{(j+1)}\in\mathbb{R}^{M\times D_{\mathrm{LWM}}}$. After passing through $L_{\mathrm{enc}}$ encoder blocks, ExtLWM produces the final CSI representation as $\mathbf{X}_{\mathrm{out},k}=\mathbf{X}_{\mathrm{enc},k}^{(L_{\mathrm{enc}}+1)}$.

\subsubsection{Masked Modeling Head}
ExtLWM retains a masked modeling head (MMH), which leverages the high-dimensional latent embeddings extracted by the preceding modules to recover the CSI. It is worth noting that, in \cite{Arxiv Alikhani}, MMH is employed only during pre-training to endow LWM-1.1 with task-agnostic and generalizable feature extraction capability. In contrast, in our framework, MMH is not only used for feature learning during training, but also continuously performs CSI extrapolation for subsequent downstream tasks. Specifically, MMH employs an FFN to map the output embedding $\mathbf{X}_{\mathrm{out},k}$ to the extrapolated CSI. Denoting the $i$-th row of $\mathbf{X}_{\mathrm{out},k}$ by $\mathbf{x}_{\mathrm{out},i,k}$, the corresponding output is given by
\begin{equation}
\label{eq_mmh_lwm}
\mathbf{x}_{\mathrm{ext},i,k} = \operatorname{FFN}(\mathbf{x}_{\mathrm{out},i,k}),  \; \mathrm{s.t.} \; [\mathbf{o}]_i=0, \forall i \in \mathcal{M},
\end{equation}
and predicts the CSI associated with the $i$-th candidate position, i.e., $i$-th row of $\mathbf{H}_{\mathrm{ext},k}$.

\vspace{-0.2cm}

\subsection{Connector}
The connector is designed to bridge the modality gap by mapping from the embeddings of ExtLWM into the linguistic embedding space of LLM. Inspired by \cite{ICLR Li}, the connector employs a lightweight Q-Former architecture for efficient cross-modal alignment. With trainable parameters $\bm{\theta}_{\mathrm{Con}}$, the connector consists of an embedding layer, a query layer, and a projection layer, which are detailed in the following.

\subsubsection{Embedding Layer}\label{section:3-B-1}
To efficiently extract semantic information from the CSI embeddings, we employ a learnable embedding matrix to generate $D_{\mathrm{q}}$ query tokens. The initialized tokens $\mathbf{Y}_{\mathrm{q}} \in\mathbb{R}^{D_{\mathrm{q}} \times D_{\mathrm{LWM}}}$ will serve as interface for cross-modal interaction to capture the latent manifold structure of the CSI embeddings.

\subsubsection{Query Layer}
The query layer is composed of $L_{\mathrm{q}}$ transformer blocks to distill language-relevant representations from CSI features. Specifically, each block consists of the following three components:
\begin{itemize}
\item Self-Attention: For the $j$-th block, an $H_{\mathrm{q}}$-head self-attention module is first applied to the query embeddings $\mathbf{Y}_{\mathrm{q}, k}^{(j)}$ to facilitate interactions among the query tokens and enhance representational diversity.

\item Cross-Attention: Subsequently, the queries $\mathbf{Y}_{\mathrm{q},k}^{(j)}$ act as ``probes'' to interact with the CSI features through an $H_{\mathrm{q}}$-head cross-attention module, where the keys and values are projected from $\mathbf{X}_{\mathrm{out},k}$. This enables the queries to selectively extract informative task-relevant features from the CSI embeddings.

\item FFN: Finally, the updated query tokens are passed through an FFN to produce the output for the $j$-th block, denoted as $\mathbf{Y}_{{\mathrm{q}},k}^{(j+1)}$.

\end{itemize}

\subsubsection{Projection Layer}
To ensure compatibility with the downstream LLM, the output of the final query block, $\mathbf{Y}_{\mathrm{q},k}^{(L_{\mathrm{q}}+1)}$, is linearly projected into the GPT input embedding space, yielding $\mathbf{Y}_{\mathrm{out},k} \in \mathbb{R}^{D_{\mathrm{q}} \times D_{\mathrm{LLM}}}$, where $D_{\mathrm{LLM}}$ denotes the hidden dimension of GPT. The projected embeddings serve as ``soft prompts'', conditioning GPT on the CSI features extracted by ExtLWM.

\vspace{-0.2cm}

\subsection{GPT-Based LLM}\label{section:3-C}
We employ a frozen LLM to unify multiple downstream wireless tasks as token-level optimization problems. In particular, for MA positioning, we exploit the pre-trained autoregressive sequential generation capability of the LLM to enable parameter-efficient adaptation. Specifically, the LLM-based cross-modal generation process can be formulated as
\begin{equation}
\label{eq_gen_llm}
t_{i+1} =  \mathrm{LLM} \big( \mathbf{Z}_{\mathrm{csi}}, \{t_{j}\}_{j=1}^i ; \bm{\theta}_{\mathrm{LLM}} \big),
\end{equation}
where $\bm{\theta}_{\mathrm{LLM}}$ denotes the pre-trained LLM parameters, $\mathbf{Z}_{\mathrm{csi}} \in \mathbb{R}^{D_{\mathrm{x}} \times D_{\mathrm{LLM}}}$ denotes the aligned CSI embeddings, and $\{t_{j}\}_{j=1}^i$ denotes the ordered sequence of $i$ generated token IDs, with $\{t_{j}\}_{j=1}^0 \triangleq \emptyset$. Specifically, for LoS/NLoS identification, $\mathbf{Z}_{\mathrm{csi}}=\mathbf{Y}_{\mathrm{out},k}$ and $D_{\mathrm{x}}=D_{\mathrm{q}}$. For sum-rate maximization, $\mathbf{Z}_{\mathrm{csi}}=\mathrm{Concat}\big(\{\mathbf{Y}_{\mathrm{out},k}\}_{k=1}^{K}\big)$ and $D_{\mathrm{x}}=KD_{\mathrm{q}}$. In this paper, we adopt the GPT-2 architecture~\cite{OPENAI Radford} as the LLM backbone due to its moderate parameter scale and practical implementation feasibility. Its main components are described as follows.

\subsubsection{Embedding Layer}

The embedding layer fuses the CSI embeddings with the textual embeddings. Specifically, GPT-2 employs the standard byte-pair encoding tokenizer and maps text tokens into $D_{\mathrm{LLM}}$-dimensional embeddings through the word-token embedding (WTE) $\mathbf{W}_{\mathrm{TE}} \in \mathbb{R}^{50257 \times D_{\mathrm{LLM}}}$ and word-position embedding $\mathbf{W}_{\mathrm{PE}} \in \mathbb{R}^{1024 \times D_{\mathrm{LLM}}}$ , which are then combined with the CSI embeddings to form the multimodal input:
\begin{equation}
\label{eq_llm_emb}
\mathbf{z}_{\mathrm{text},i}^T=[\mathbf{W}_{\mathrm{TE}}]_{t_i,:} + [\mathbf{W}_{\mathrm{PE}}]_{i,:},
\end{equation}
where $\mathbf{z}_{\mathrm{text}, i}  \in \mathbb{R}^{D_{\mathrm{LLM}}}$ denotes the text embedding of the $i$-th word token, and $t_{i} \in \{0,1,...,50256\}$. Packing all textual embedding into the matrix $\mathbf{Z}_{\mathrm{text}}^T=[\mathbf{z}_{\mathrm{text},1},...,\mathbf{z}_{\mathrm{text}, i}]$, the final multimodal input sequence $\mathbf{Z}_{\mathrm{emb}}$ is constructed by prepending the CSI embeddings to the textual prefix
\begin{equation}
\label{eq_multi_emb}
\mathbf{Z}_{\mathrm{emb}} = \left[\begin{array}{c}
\mathbf{Z}_{\mathrm{csi}} \\
\mathbf{Z}_{\mathrm{text}}
\end{array}\right] \in \mathbb{R}^{(D_{\mathrm{x}}+ i ) \times D_{\mathrm{LLM}}}. \\
\end{equation}

\subsubsection{Decoder Layer}
The backbone adopts a decoder-only architecture consisting of $L_{\mathrm{dec}}$ cascaded decoder blocks with identical structures, which are employed to progressively extract hierarchical latent representations. The $j$-th decoder block takes $\mathbf{Z}_{\mathrm{dec}}^{(j)}$ as the input and outputs the enhanced representation $\mathbf{Z}_{\mathrm{dec}}^{(j+1)}$, where the initial input is given by $\mathbf{Z}_{\mathrm{dec}}^{(1)}=\mathbf{Z}_{\mathrm{emb}}$. Specifically, within each block, through normalizing and linearly projecting $\mathbf{Z}_{\mathrm{dec}}^{(j)}$, query, key and value matrices $\mathbf{Q}_{\mathrm{dec}}^{(j)}$, $\mathbf{K}_{\mathrm{dec}}^{(j)}$, and $\mathbf{V}_{\mathrm{dec}}^{(j)} \in\mathbb{R}^{(D_{\mathrm{x}}+ i)\times D_{\mathrm{LLM}}}$ are obtained and then split into $H_{\mathrm{dec}}$ heads to compute multi-head self-attention, with the $h$-th head output given as
\begin{equation}
\label{eq_mha_lwm}
\hspace{-0.1cm} \mathbf{Z}_{\mathrm{dec}, h}^{(j)} = \operatorname{Softmax} \Bigg(\frac{\mathbf{Q}_{\mathrm{dec}, h}^{(j)}\big( \mathbf{K}_{\mathrm{dec}, h}^{(j)} \big)^T}{\sqrt{ \frac{D_{\mathrm{LLM}}}{H_{\mathrm{dec}}}}} + \mathbf{M}_{\mathrm{dec}} \Bigg)\mathbf{V}_{\mathrm{dec}, h}^{(j)},
\end{equation}
where $\mathbf{M}_{\mathrm{dec}} \in\mathbb{R}^{(D_{\mathrm{x}}+ i)\times (D_{\mathrm{x}}+ i)}$ is the multimodal causal mask. The mask $\mathbf{M}_{\mathrm{dec}}$ enables bidirectional attention among the $D_{\mathrm{x}}$ CSI tokens while preserving causal attention for the textual tokens. The outputs $\mathbf{Z}_{\mathrm{dec}, h}^{(j)}$ from all attention heads are then concatenated and processed by the FFN to obtain $\mathbf{Z}_{\mathrm{dec}}^{(j+1)}$.

\subsubsection{Output Layer}\label{section:3-C-3}
The output layer maps the final decoder representation $\mathbf{Z}_{\mathrm{dec}}^{(L_{\mathrm{dec}}+1)}$ to vocabulary logits for next-token generation. Following GPT-2, the output projection shares the token embedding matrix $\mathbf{W}_{\mathrm{TE}}$, reducing the parameter count while maintaining consistency with the pre-trained token space. Specifically, the unnormalized logits for the $(i+1)$-th token are given by:
\begin{equation}
\label{eq_llm_logits}
\mathbf{l}_{\mathrm{text}, i+1} =  \mathbf{W}_{\mathrm{TE}} \big( [\mathbf{Z}_{\mathrm{dec}}^{(L_{\mathrm{dec}}+1)}]_{-1,:} \big)^T.
\end{equation}

The generation of the $(i+1)$-th token is subsequently executed by selecting the token with the highest posterior probability:
\begin{equation}
\label{eq_llm_gen_token}
t_{i+1}=\arg \max_{t \in \mathcal{T}_{\mathrm{x}}} \bigg[ \operatorname{Softmax} \Big( \big\{ [\mathbf{l}_{\mathrm{text}, i+1}]_{t'} \big\}_{t' \in \mathcal{T}_{\mathrm{x}}} \Big) \bigg]_t,
\end{equation}
where $\mathcal{T}_{\mathrm{x}}$ denotes a task-specific restricted subset of tokens tailored for $\mathrm{x} \in \{ \mathrm{id}, \mathrm{max}\}$.
\begin{itemize}
\item For LoS/NLoS identification, $\mathcal{T}_{\mathrm{id}} = \{ 8505, 3919 \}$, corresponding to word unit ``$\mathrm{yes}$'' and ``$\mathrm{no}$'', respectively, which enables GPT-2 to leverage its inherent linguistic priors to output classification results in a semantic manner.

\item For sum-rate maximization, $\mathcal{T}_{\mathrm{max}}$ is defined as the set of token IDs corresponding to the position indices $\{\text{``}m\text{''}\}_{m=1}^{M}$. By leveraging numerical tokens with explicit stable semantics, the discrete position optimization of $N$ MAs can be naturally transformed into a sequential generation problem involving $N$ consecutive tokens\footnote{Specifically, GPT-2 adopts the ``r50k\_base'' tokenizer, where the consecutive integers from $0$ to $520$ can be encoded as single tokens. Given the candidate-position ranges commonly considered in existing MA studies~\cite{JSTSP Shao},~\cite{TWC Wang},~\cite{CL Cang}, this range is sufficient for the settings considered in this work. For larger candidate-position spaces beyond the single-token range, we design a multi-token position representation and generation mechanism, as demonstrated in Fig.~\ref{fig_max_LLM}.}.
\end{itemize}

\section{Multi-Stage Training Strategy}\label{section:4}
Building upon the unified framework, this section develops a complete multi-stage training algorithm to enable parameter-efficient optimization of CSI extrapolation, LoS/NLoS identification, and sum-rate maximization under data-limited conditions. Details of training stages will be specified in the following subsections.

\vspace{-0.2cm}
\subsection{Stage-1: Pre-Training for ExtLWM}
The pre-training stage encourages ExtLWM to learn generic structural features of MA channels from a large number of diverse and complex scenarios. Its optimization objective is to train the MMH to reconstruct the original unmasked CSI in \eqref{eq_patch} based on the output in \eqref{eq_mmh_lwm}. Accordingly, the self-supervised training loss is formulated as:
\begin{equation}
\label{eq_loss_pre}
 \hspace{-0.08cm} L_{\mathrm{pre}}(\bm{\theta}_{\mathrm{WFM}})= \mathbb{E}_{k \in \mathcal{K}_{\mathrm{pre}}} \Bigg[ \sum_{i\in \mathcal{I}_k, o_{i}=0} \frac{\| \mathbf{x}_{\mathrm{ext}, i, k} - \mathbf{x}_{i,k} \|^2}{\| \mathbf{x}_{i,k} \|^2} \Bigg],
\end{equation}
where $\mathcal{K}_{\mathrm{pre}}$ denotes the set of users used for pre-training, $\mathcal{I}_k$ denotes the set of  masked patch indices (representing candidate antenna positions) for the $k$-th user, with its cardinality varying across different pre-training scenarios.

To further enhance the CSI extrapolation capability of ExtLWM  under highly incomplete CSI, we increase the masking ratio from about $40\%$ in LWM-1.1 to $80\%$, exposing the model to more severe CSI incompleteness during pre-training. Meanwhile, the masking indices $\mathcal{I}_k$ are randomized in each epoch, encouraging ExtLWM to learn spatial priors that are more robust to different CSI observation patterns and thereby improving its ability to capture informative channel features from sparse CSI.

\vspace{-0.2cm}
\subsection{Stage-2: Fine-Tuning for CSI Extrapolation}
Building on the paradigm of large-scale pre-training followed by task-specific fine-tuning, we perform full-parameter fine-tuning of $\bm{\theta}_{\mathrm{WFM}}$ to adapt the generalized spatial representations to scenario-specific propagation characteristics. This process is particularly effective in data-scarce regimes, where ExtLWM leverages the learned channel priors to achieve high-fidelity CSI extrapolation with limited labeled samples.

Considering \eqref{eq_problem}, stage-2 aims to fully exploit the CSI extrapolation capability of ExtLWM to satisfy~\eqref{eq_constraint_nmse}, and its training objective is formulated as
\begin{equation}
\label{eq_min_mmse}
\min_{\bm{\theta}_{\mathrm{WFM}}}  ~ \mathbb{E}_{k \in \mathcal{K}} \big[ \mathrm{NMSE}_k \big].
\end{equation}

Meanwhile, the training loss for stage-2 is formulated as
\begin{equation}
\label{eq_loss_ext}
 \hspace{-0.08cm} L_{\mathrm{ext}}(\bm{\theta}_{\mathrm{WFM}})= \mathbb{E}_{k \in \mathcal{K}_{\mathrm{ext}}} \Bigg[ \sum_{i\in \mathcal{M}, o_{i}=0} \frac{\| \mathbf{x}_{\mathrm{ext}, i, k} - \mathbf{x}_{i,k} \|^2}{\| \mathbf{x}_{i,k} \|^2} \Bigg],
\end{equation}
where $\mathcal{K}_{\mathrm{ext}}$ denotes the set of users used for fine-tuning with complete CSI. Upon completion of stage-2, we will freeze $\bm{\theta}_{\mathrm{WFM}}$ to preserve its channel feature extraction and CSI extrapolation capabilities.

\vspace{-0.2cm}
\subsection{Stage-3: Fine-Tuning for LoS/NLoS Identification}
The fine-tuned ExtLWM can effectively extract and output scenario-specific channel feature embeddings, serving as a shared feature backbone for multiple downstream wireless tasks. Based on \eqref{eq_constraint_f1}, stage-3 aims to optimize the connector to enable the adaptation of LLM to LoS/NLoS identification. The training objective of stage-3 can be expressed as
\begin{equation}
\label{eq_max_F1}
\max_{\bm{\theta}_{\mathrm{Con}}}  ~  \mathrm{F1}\text{-}\mathrm{score}(\mathcal{K}),
\end{equation}and the training loss of stage-3 can be formulated as
\begin{equation}
\label{eq_loss_cls}
 L_{\mathrm{id}}(\bm{\theta}_{\mathrm{Con}})= \mathbb{E}_{k \in \mathcal{K}_{\mathrm{id}}} \big[ -[\delta_k \log p_k^{1}+(1-\delta_k) \log p_k^{0}] \big],
\end{equation}
where $\mathcal{K}_{\mathrm{id}}$ denotes the set of users used for training, for which only incomplete CSI samples and the corresponding LoS/NLoS labels are required, $\delta_k \in \{0, 1\}$ is the ground-truth label and $p_k^i = \text{Pr}(\hat{\delta}_k = i)$ is the predicted probability for class $i$. Recall \eqref{eq_llm_gen_token}, $p_k^1$ and $p_k^0$ can also be interpreted as the softmax-normalized scores for the semantic tokens, i.e., ``$\mathrm{yes}$'' and ``$\mathrm{no}$'', respectively.

\vspace{-0.2cm}
\subsection{Stage-4: RL for Sum-Rate Maximization}
In this subsection, we proceed to maximize sum-rate formulated in \eqref{eq_max_rate}. Distinctively, we propose a paradigm shift by analogizing the antenna positioning to token prediction in natural language processing. Just as an LLM predicts the most probable ``optimal'' tokens within a massive vocabulary based on linguistic context, we aim to identify the optimal antenna positions from a large number of candidates. The core philosophy is to leverage the feature extraction capability of the LLM to interpret wireless channels. By treating the previously selected antenna positions and their corresponding channel states as context, the model is empowered to predict the subsequent optimal positions.

Subsequently, we formulate MA positioning as an MDP to capture sequential dependencies and align it with LLM autoregressive generation. The key elements of the MDP are defined as follows:

\begin{itemize}
\item State $\mathsf{s}_n$: $\mathsf{s}_n$ encapsulates the multimodal context, comprising the CSI embeddings output by ExtLWM and the previously generated text tokens:
\begin{equation}
\mathsf{s}_n \triangleq \big\{ \big\{ \mathbf{X}_{\mathrm{out}, k} \big\}_{k=1}^K, \{t_{j}\}_{j=1}^{n-1} \big\}, \forall n \in \mathcal{N}.
\end{equation}
\item Action $\mathsf{a}_n$: $\mathsf{a}_n$ represents the prediction of $n$-th token ID $t_n$. Recall that we establish a semantic mapping between the discrete space  $\mathcal{T}_{\mathrm{max}}$ and the physical coordinate $p_n$ of the $n$-th MA, ensuring \eqref{eq_constraint_p} is satisfied.

\item Action mask $\mathsf{m}_n$: We incorporate a mask vector $\mathsf{m}_n \in \{0, 1\}^{M}$ to prevent LLM from generating invalid $\mathbf{p}$ violating the minimum antenna spacing constraint \eqref{eq_constraint_pos}. Specifically, $m_n$ denotes the mask over $\mathcal{T}_{\mathrm{max}}$ that identifies the valid candidate tokens for $t_n$, i.e., the feasible positions of the $n$-th MA:
\begin{equation}
\label{eq_action_mask}
\hspace{-0.3cm} \mathsf{m}_n \in \{0, 1\}^{M}, \big[\mathsf{m}_n \big]_i = \begin{cases} 1, i \in \big[M_{\min , n}, M_{\max , n}\big] \\ 0,  \mathrm{else} \end{cases},
\end{equation}
where $M_{\min,n} = 1$ for $n=1$ and $M_{\min,n} = p_{\,n-1} + \frac{d_{\min}}{D}$ for $n>1$, while $M_{\max,n} = M - (N-n)\frac{d_{\min}}{D}$. After the masking process, the logits corresponding to all invalid tokens are set to $-\infty$.

\item Policy $\pi$: The policy $\pi(\cdot \mid \mathsf{s}_n, \mathsf{m}_n)$ depicts a probability distribution over the candidate set $\mathcal{T}_{\max}$ with the multimodal context and physical constraints given. In the training phase, the policy gradient algorithm iteratively refines $\pi(\mathsf{a}_n| \mathsf{s}_n, \mathsf{m}_n)$ , enabling the model to learn the mapping to generate high-sum-rate antenna positions.

\item Reward $\mathsf{r}$: A scalar reward is computed after generating all $N$ actions. In data-limited scenarios, the ground-truth CSI of all users may not be available. Fortunately, the reward constructed from the high-quality extrapolated CSI provided by ExtLWM can still effectively guide policy optimization, and is defined as:
\begin{equation}
\label{eq_mdp_reward}
\mathsf{r} \triangleq \sum_{s=1}^S \sum_{k=1}^{K}\log_2 \left(1 + \frac{|\mathbf{\overline{h}}_{\mathrm{ext},k,s}^H \mathbf{w}_{k, s} |^2}{\sum_{j \neq k} |\mathbf{\overline{h}}_{\mathrm{ext},k, s}^H \mathbf{w}_{j, s} |^2+\sigma_{\mathrm{c}}^2}\right),
\end{equation}
where $\mathbf{w}_{k,s}$ is the $k$-th column of $\mathbf{W}_s$. Various precoding schemes can be seamlessly incorporated to jointly construct the reward function, such as MMSE, weighted MMSE (WMMSE)~\cite{TSP Shi}, and zero forcing (ZF). In this paper, we adopt MMSE precoder to achieve a favorable trade-off between sum-rate performance and computational efficiency. Specifically, assuming that the total transmit power is equally allocated across all subcarriers, the MMSE precoder can be expressed as
\begin{equation}
\label{eq_MMSE}
\mathbf{W}_s = \frac{\sqrt{P_{s}}\mathbf{\overline{H}}_{\mathrm{ext}, s} (P_{s} \mathbf{\overline{H}}_{\mathrm{ext}, s}^H \mathbf{\overline{H}}_{\mathrm{ext}, s}+\sigma_{\mathrm{c}}^2 \mathbf{I}_K)^{-1}}{\| \mathbf{\overline{H}}_{\mathrm{ext}, s} (P_{s} \mathbf{\overline{H}}_{\mathrm{ext}, s}^H \mathbf{\overline{H}}_{\mathrm{ext}, s}+\sigma_{\mathrm{c}}^2 \mathbf{I}_K)^{-1} \|_F},
\end{equation}
where $P_{s}=\frac{P_{\mathrm{tx}}}{S}$ and  $\mathbf{\overline{H}}_{\mathrm{ext}, s} \triangleq [\mathbf{\overline{h}}_{\mathrm{ext},1, s},..., \mathbf{\overline{h}}_{\mathrm{ext},K, s}]$.
\end{itemize}

The constructed MDP provides a natural framework for deploying RL methods to optimize the sequential MA positioning policy. To efficiently preserve the basic semantic and LoS/NLoS classification capabilities of the LLM, we only train $\bm{\theta}_{\mathrm{Con}}$ while keeping the other modules frozen. Therefore, in stage-4, the original problem~\eqref{eq_problem} can be reformulated as
\begin{subequations}
\label{eq_mdp_problem}
\begin{align}
\max _{\bm{\theta}_{\mathrm{Con}}}  & \, \, \, \mathbb{E}_{\tau\sim\pi }\big[\mathsf{r}(\tau)\big], \label{eq_max_reward}  \\
\mathrm{ s.t. } & \,\,\, \mathrm{F1}\text{-}\mathrm{score}(\mathcal{K}) \geq \epsilon_{\mathrm{F1}},  \label{eq_constraint_cl}
\end{align}
\end{subequations}
where $\tau \triangleq \big\{(\mathsf{s}_n, \mathsf{a}_n, \mathsf{m}_n) \big\}_{n=1}^{N}$ denotes the roll-out trajectory. $\mathsf{r}(\cdot)$ denotes the trajectory reward specified in \eqref{eq_mdp_reward}. Note that the constraints in \eqref{eq_constraint_pos} and \eqref{eq_constraint_p} are guaranteed by the action space and masking mechanism, respectively, while \eqref{eq_constraint_power} is satisfied by the normalized precoder and \eqref{eq_constraint_nmse} is addressed in stage-2. Hence, only \eqref{eq_constraint_f1}, i.e., \eqref{eq_constraint_cl}, needs to be explicitly handled. For each training epoch, we denote the $i$-th trajectory as $\tau^{(i)}$, and define its $n$-th transition tuple as $\mathsf{b}_n^{(i)} \triangleq \big(\mathsf{s}_n^{(i)}, \mathsf{a}_n^{(i)}, \mathsf{m}_n^{(i)}, \pi(\mathsf{a}_n^{(i)}| \mathsf{s}_n^{(i)}, \mathsf{m}_n^{(i)}), \mathsf{s}_{n+1}^{(i)}, \mathsf{r}^{(i)} \big)$, which is sequentially stored into buffer $\mathcal{B}$.

To fully exploit the potential of the LLM for sequential optimization, we adopt the GRPO-based RL algorithm, an LLM-specific and training-efficient method that does not require an additional critic network~\cite{Arxiv Shao}. Specifically, GRPO directly performs advantage estimation based on group-relative rewards, thereby effectively reducing the variance of advantage estimation and alleviating the credit-assignment difficulty in sequential decision making. The advantage value of the $i$-th trajectory is computed based on the $G$ trajectories rolled out in the current training epoch:
\begin{equation}
\label{eq_grpo_adv}
A^{(i)}= \frac{\mathsf{r}^{(i)} - \operatorname{mean}( \{\mathsf{r}^{(j)}\}_{j=1}^G )}{\operatorname{std}( \{\mathsf{r}^{(j)}\}_{j=1}^G )}.
\end{equation}where $r^{(i)}$ is the terminal reward of the $i$-th trajectory, which is inherently delayed and sparse since the marginal contribution of each MA cannot be explicitly measured.
  The GRPO loss $L_{\mathrm{rl}}$ is then formulated using the clipped surrogate objective to constrain the update magnitude:
\begin{equation}
\label{eq_loss_act}
L_{\mathrm{rl}} = \mathbb{E}_{ \mathcal{B}_{\mathrm{sub}}} \left[ \min \left( \rho_n^{(i)} A^{(i)}, \mathrm{clip}(\rho_n^{(i)}, 1-\epsilon_{\mathrm{c}}, 1+\epsilon_{\mathrm{c}}) A^{(i)} \right) \right],
\end{equation}
where $\mathcal{B}_{\mathrm{sub}}$ denotes the mini-batch subsets randomly drawn from $\mathcal{B}$, $\rho_n^{(i)}= \frac{\pi_{\mathrm{new}}(\mathsf{a}_{n}^{(i)} \mid \mathsf{s}_{n}^{(i)}, \mathsf{m}_{n}^{(i)})}{\pi (\mathsf{a}_{n}^{(i)} \mid \mathsf{s}_{n}^{(i)}, \mathsf{m}_{n}^{(i)})}$ represents the probability ratio between the updated and old policies.

Meanwhile, to improve the training stability of GRPO, we introduce entropy regularization to better balance exploration and exploitation. The exploration loss can be expressed as
\begin{equation}
\label{eq_loss_exp}
\hspace{-0.05cm} L_{\mathrm{exp}} = \mathbb{E}_{\mathcal{B}_{\mathrm{sub}}} \Big[ \mathcal{H}\big( \pi_{\mathrm{new}}( \cdot ~| \mathsf{s}_{n}^{(i)}, \mathsf{m}_{n}^{(i)}) \big) \Big],
\end{equation}
where $\mathcal{H}(\cdot)$ denotes the policy entropy.

Moreover, to preserve the LoS/NLoS classification performance in \eqref{eq_constraint_cl} during sum-rate maximization, we incorporate experience replay (ER), a widely used strategy in continual learning (CL) \cite{NeurIPS Rolnick}, to mitigate catastrophic forgetting.  Specifically, ER can be naturally integrated with the replay buffer $\mathcal{B}$ of the underlying RL algorithm by additionally storing a fixed set of labeled samples $(\delta_k,\mathbf{H}_{\mathrm{inc},k})$, which provides stable supervision throughout training and regularizes policy updates. Since only a small subset $\mathcal{K}_{\mathrm{cl}}\subseteq\mathcal{K}_{\mathrm{id}}$ is required to effectively preserve the classification capability, the additional memory overhead introduced by ER remains limited. Finally, within each epoch, the loss function for sum-rate maximization can be expressed as 
\begin{equation}
\label{eq_loss_max}
L_{\mathrm{max}}(\bm{\theta}_{\mathrm{Con}}) = -L_{\mathrm{rl}} - \lambda_{\mathrm{exp}} L_{\mathrm{exp}} + \lambda_{\mathrm{cl}} L_{\mathrm{cl}},
\end{equation}
where $ L_{\mathrm{cl}}= \mathbb{E}_{k \in \mathcal{K}_{\mathrm{cl}}} \big[ -[\delta_k \log p_k^{1}+(1-\delta_k) \log p_k^{0}] \big]$  is defined similarly to \eqref{eq_loss_cls},  $\lambda_{\mathrm{exp}}$ and $\lambda_{\mathrm{cl}}$ are hyperparameters controlling the priority of exploration and ER strategy, respectively.

\vspace{-0.2cm}
\subsection{Summary}

We obtain ExtLWM through pre-training in stage-1. For a specific downstream MA system, stages-2-4 progressively endow the multimodal model with multi-task capabilities under data-limited conditions to solve problem~\eqref{eq_problem}. Notably, after stage-2, ExtLWM serves as an environment-specialized model to support a \textbf{self-contained} framework, providing both high-quality feature embeddings and extrapolated CSI to support the token-level optimization in stages-3/4 and the construction of the reward function (including precoder) in stage-4, respectively.

We now analyze the computational complexity of the proposed multimodal model for generating $N$ MA positions. Specifically, the incomplete CSI of the $K$ users is first processed by ExtLWM. The CSI of each user is successively passed through the embedding layer and $L_{\mathrm{enc}}$ encoder blocks, resulting in a computational complexity of
$\mathcal{O}\big(KMSD_{\mathrm{LWM}}+$\\$KL_{\mathrm{enc}}(M^2D_{\mathrm{LWM}}+ MD_{\mathrm{LWM}}^2)\big)$. The extracted CSI embeddings are then fed into the connector, where $D_\mathrm{q}$ learnable query tokens and $L_{\mathrm{q}}$ query blocks are employed to project the CSI features into the embedding space of the LLM, leading to a computational complexity of $\mathcal{O}\big(KL_{\mathrm{q}}(D_\mathrm{q}^2D_{\mathrm{LWM}}+MD_\mathrm{q}$\\$D_{\mathrm{LWM}}+(M+D_\mathrm{q})D_{\mathrm{LWM}}^2)+KD_\mathrm{q}D_{\mathrm{LWM}}D_{\mathrm{LLM}}\big)$. Subsequently, GPT-2 autoregressively generates the $N$ MA position tokens based on the resulting CSI embeddings, with a computational complexity of
$\mathcal{O}\big(L_{\mathrm{dec}}(D_\mathrm{x}^2D_{\mathrm{LLM}}+ D_\mathrm{x}D_{\mathrm{LLM}}^2)+L_{\mathrm{dec}}$\\$(ND_{\mathrm{LLM}}^2+(ND_\mathrm{x}+N^2) D_{\mathrm{LLM}})+NMD_{\mathrm{LLM}}\big)$, where $D_\mathrm{x}=KD_\mathrm{q}$. Therefore, by neglecting lower-order terms and constant factors, the dominant computational complexity of the proposed model can be expressed as
$\mathcal{O}\big(KL_{\mathrm{enc}}(M^2D_{\mathrm{LWM}}+$\\$MD_{\mathrm{LWM}}^2)+KL_{\mathrm{q}}MD_{\mathrm{LWM}}^2+L_{\mathrm{dec}}(ND_{\mathrm{LLM}}^2+N^2D_{\mathrm{LLM}})+NMD_{\mathrm{LLM}}\big)$.

\vspace{-0.2cm}

\section{Numerical Results}\label{section:5}
In this section, we provide comprehensive simulation results and performance analysis. We first evaluate the pre-training and CSI extrapolation performance of ExtLWM. Next, we evaluate the LoS/NLoS identification capability of the proposed network. Finally, we evaluate the LLM-based MA positioning scheme in terms of sum-rate performance. All experiments were conducted on a hardware platform equipped with two NVIDIA GeForce RTX 4090 GPUs.

\vspace{-0.2cm}

\subsection{Pre-Training Performance}
Consistent with LWM-1.1~\cite{Arxiv Alikhani}, the pre-training dataset comprises 120 scenarios from DeepMIMO-V3~\cite{ITA Alkhateeb} with $f_{\mathrm{c}}=3.5$ GHz and $\Delta f=30$ kHz, while the ExtLWM (2.4M parameters) hyperparameters are set to $\alpha_{\mathrm{p}}=10^6$, $[\mathrm{MASK}]=0.1$, $D_{\mathrm{LWM}}=128$, $L_{\mathrm{enc}}=12$, and $H_{\mathrm{enc}}=8$. Among all users, $75\%$ of the CSI samples---corresponding to $|\mathcal{K}_{\mathrm{pre}}|=402409$ ---are used for training. In this paper, we train three WFMs to accommodate downstream MA systems with $D=\frac{\lambda_{\mathrm{c}}}{8}$, as summarized in Table \ref{table_pre_set}. In terms of pre-training configuration, while retaining most settings from LWM-1.1, we set epoch $E_{\mathrm{pre}}=800$ to more effectively capture the latent structure of CSI. The peak GPU memory consumption during pre-training is approximately $23.7$GB.

\begin{table}[t]
    \centering
    \footnotesize
    \caption{Training configurations for three WFMs.}
    \label{table_pre_set}
    \begin{tabular}{|c|c|c|c|}
    \hline
    Model Name   & Pre-training candidate position $| \mathcal{I}_k |$ & $S$ & Cost \\
    \hline
    ExtLWM-sub16 & 40 80 120 160 200 240 280 320  & 16 & 98h\\
    \hline
    ExtLWM-sub32 & 30 60 90 120 160 & 32 & 78h\\
    \hline
    ExtLWM-sub64 & 20 40 80 & 64 & 76h\\
    \hline
    \end{tabular}
    \vspace{-0.2 cm}

\end{table}

\begin{figure}[t]
\centering
\includegraphics[width = 3.0 in]{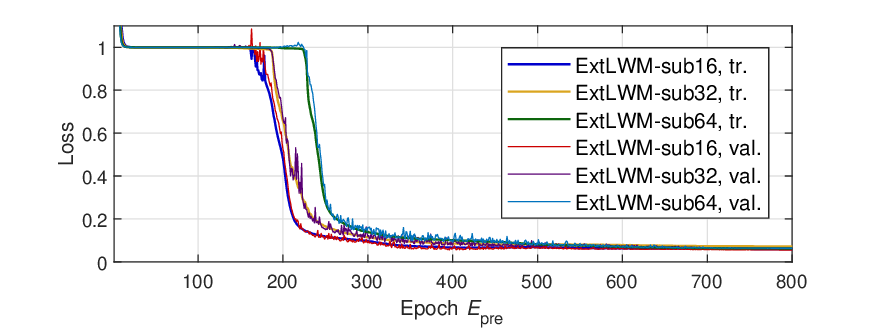}
\caption{Training and validation losses for WFMs.}
\label{fig_pre_S}
\vspace{-0.3cm}
\end{figure}

\begin{figure}[t]
\centering
\includegraphics[width = 3.0 in]{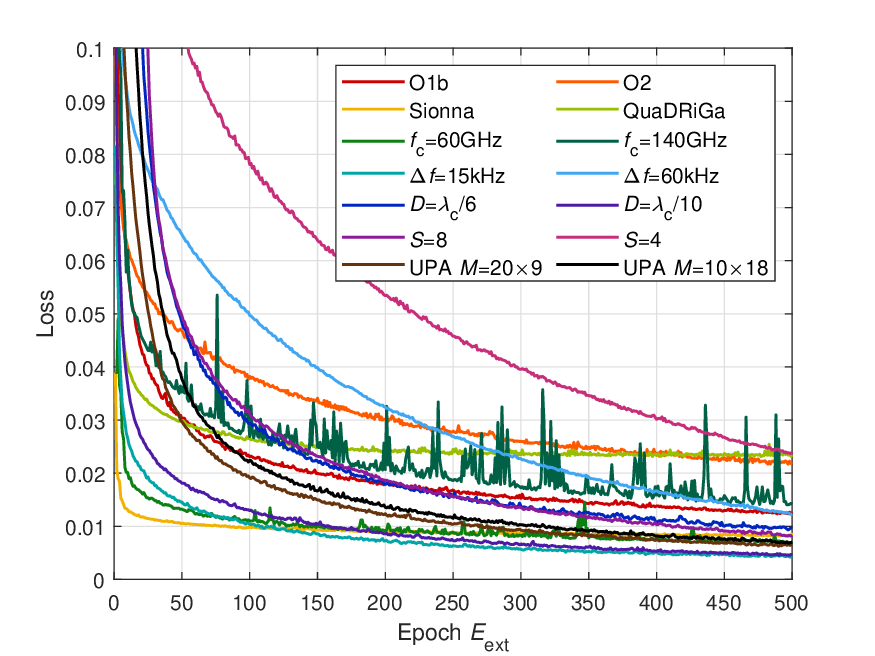}
\caption{ExtLWM's performance under various conditions.}
\label{fig_pre_scen}
\vspace{-0.3 cm}
\end{figure}

Fig. \ref{fig_pre_S} presents the loss trajectories of the entire training phase of the three foundation models, demonstrating the effectiveness and stability of the training process. In this paper, we adopt ``\textbf{ExtLWM-sub16}''  as the foundation model for all subsequent experiments, referred to simply as \textbf{ExtLWM}\footnote{Available: https://huggingface.co/Login2025/ExtLWM}.

Next, we separately fine-tune ExtLWM on 14 sets of out-of-distribution CSI samples, where 50\% of the samples in each unseen MA dataset are used for training. Fig.~\ref{fig_pre_scen} presents the validation NMSE under $M=180$ and $\eta_{\mathrm{msk}}=80\%$. It can be observed that all curves rapidly decrease below $0.05$ within 500 epochs, demonstrating the generalization capability of ExtLWM across diverse unseen channel distributions from multiple perspectives. Specifically, 1)~We consider two DeepMIMO scenarios that are not included in the pre-training dataset, namely ``O1-blockage'' and ``O2-dynamic'', to demonstrate the cross-scenario generalization capability of ExtLWM. 2)~We evaluate ExtLWM using CSI generated by QuaDRiGa \cite{TAP Jaeckel} and Sionna \cite{Arxiv Aoudia}, thereby assessing its generalization capability across different channel-generation platforms beyond DeepMIMO. 3)~We evaluate channel samples generated at $f_{\mathrm{c}}=60$ GHz and $f_{\mathrm{c}}=140$ GHz to investigate the cross-frequency generalization capability of ExtLWM. 4)~We consider $\Delta f=15$ kHz and $\Delta f=60$ kHz  to evaluate its generalization across different subcarrier spacings. 5)~We further generate CSI using different candidate-position spacings, i.e., $D=\frac{\lambda_{\mathrm{c}}}{6}$ and $D=\frac{\lambda_{\mathrm{c}}}{10}$, to validate the cross-array-spacing generalization capability of ExtLWM beyond the pre-training grid with $\frac{\lambda_{\mathrm{c}}}{8}$. 6)~We evaluate channel configurations with $S=8$ and $S=4$, thereby assessing the generalization capability of ExtLWM across different numbers of subcarriers. 7)~Finally, we evaluate ExtLWM using uniform planar array (UPA) channels with $M=20\times9$ and $M=10\times18$, demonstrating that ExtLWM can capture spatial and propagation structures beyond a specific ULA geometry.

\vspace{-0.2cm}

\subsection{CSI Extrapolation Performance}
In this subsection, we demonstrate the CSI extrapolation performance of fine-tuned ExtLWM in the specific downstream scenario, which is constructed on the data of 1250 users lying between row-1400 and row-1500 in ``Boston5G'' ($M=180$, $\eta_{\mathrm{msk}}=80\%$, $f_{\mathrm{c}} = 28$ GHz, $S=16$ and $D=\frac{\lambda_{\mathrm{c}}}{8}$). To simulate the sample-limited environment, 50\% of CSI samples, i.e., $|\mathcal{K}_{\mathrm{ext}}|=625$, are used for fine-tuning ExtLWM. The batch size, training epochs, and learning rate are set to 128, $E_{\mathrm{ext}}=2 \times 10^{4}$, and $\eta_{\mathrm{ext}}=5 \times 10^{-5}$, respectively. The peak GPU memory consumption and training time for stage-2 are approximately $8.3$GB and $3.5$h, respectively.

To provide a comprehensive performance comparison, we test the following benchmarks:

\begin{itemize}
\item LWM (60 scenarios): The number of pre-training scenarios for ExtLWM is reduced from 120 to 60.
\item LWM (1 scenario): A dedicated extrapolation model is first trained on the ``ASU Campus'' scenario and then fully fine-tuned on the target scenario.

\item LWM-1.1 \cite{Arxiv Alikhani}: For the case of $S=16$, the original patch shape of $2\times4\times4$ is reconfigured to $2\times1\times16$, enabling LWM-1.1 to be directly applied to CSI extrapolation.
\item LWM-scratch: A transformer with the same architecture as LWM is trained from scratch.
\item Bi-LSTM: We replace the encoder blocks of LWM with a Bi-LSTM architecture having a comparable number of parameters and train the resulting model from scratch.
\item BERT \cite{TASLP Cui}: Following \cite{ICC Wu}, we replace the embedding layer and MMH of the BERT with the corresponding ExtLWM modules, and fine-tune the last four encoder blocks.
\end{itemize}

\begin{figure}[t]
\centering
\includegraphics[width = 3.0 in]{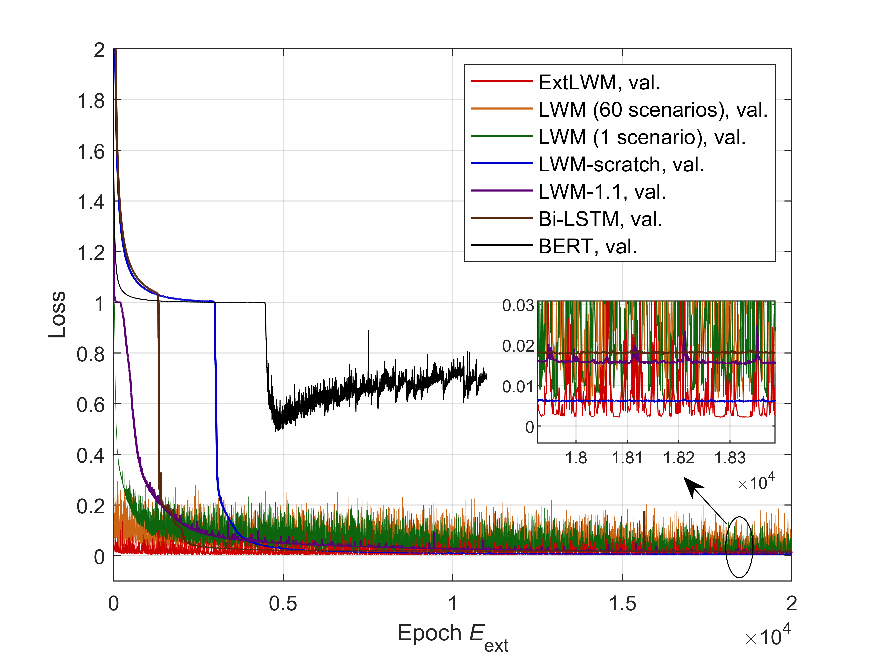}
\caption{Validation loss for CSI extrapolation.}
\label{fig_ext_loss}
\vspace{-0.3 cm}
\end{figure}

Fig.~\ref{fig_ext_loss} shows the validation loss curves of different models. It can be observed that both ExtLWM variants start from relatively low NMSE levels and then decrease rapidly during fine-tuning, indicating that MA-specific pre-training provides effective channel representations for CSI extrapolation. In comparison, LWM-1.1 remains in the region around an NMSE of 1 for a noticeably shorter period than the model trained from scratch, suggesting that its pre-training still preserves certain generic representations of geometric wireless channels. Finally, the BERT-based model consistently exhibits a higher validation loss, indicating a mismatch between its pre-trained representations and wireless channel characteristics, which cannot be sufficiently compensated for using only a limited number of downstream CSI samples.

\begin{figure}[t]
\centering
\includegraphics[width = 3.0 in]{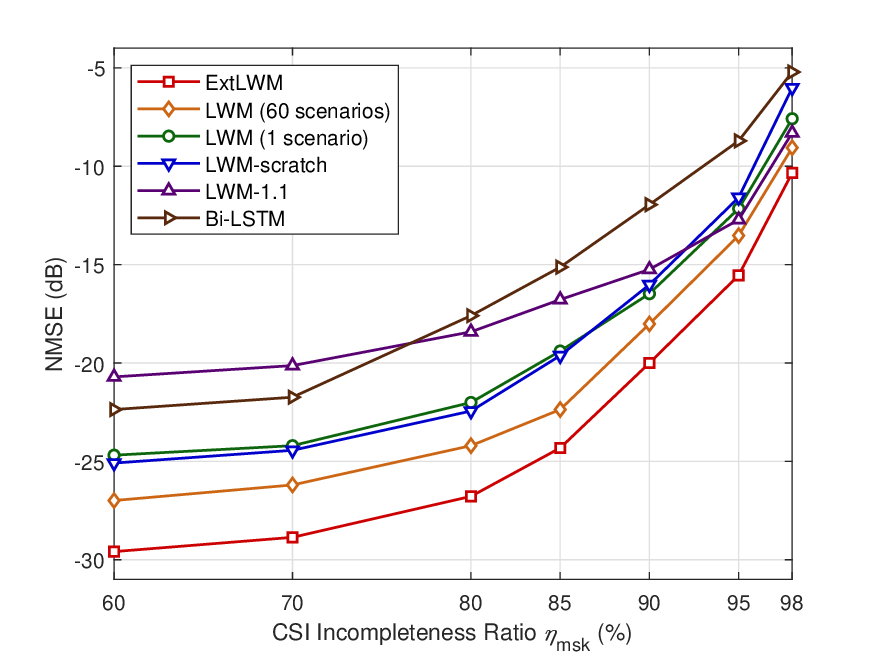}
\caption{Impact of CSI incompleteness ratio.}
\label{fig_ext_msk}
\vspace{-0.3 cm}
\end{figure}

In Fig. \ref{fig_ext_msk}, we evaluate the CSI extrapolation performance under different CSI incompleteness ratios $\eta_{\mathrm{msk}}$. The results show that ExtLWM consistently achieves superior performance. When the number of pre-training scenarios is increased from 60 to 120, ExtLWM achieves an average CSI extrapolation gain of approximately $2.15$~dB, indicating that pre-training over more diverse scenarios facilitates the learning of more transferable spatial correlations and multipath propagation characteristics. Meanwhile, compared with the model trained from scratch, the standard ExtLWM achieves NMSE gains of more than $4.5$~dB and $2.8$~dB at CSI incompleteness ratios of $60\%$ and $95\%$, respectively. Notably, even when pre-trained using only 60 scenarios, ExtLWM still outperforms the state-of-the-art LWM-1.1, further demonstrating the necessity of MA-specific WFM pre-training. Specifically, LWM-1.1 is primarily designed to learn spatial--frequency correlations from regular FPA channels and thus has limited capability in characterizing the distinctive spatial distribution patterns of MA channels. Moreover, LWM-1.1 adopts a masking ratio of only $40\%$ during pre-training, resulting in insufficient exposure to highly incomplete CSI patterns.

\begin{figure}[t]
\centering
\includegraphics[width = 3.0 in]{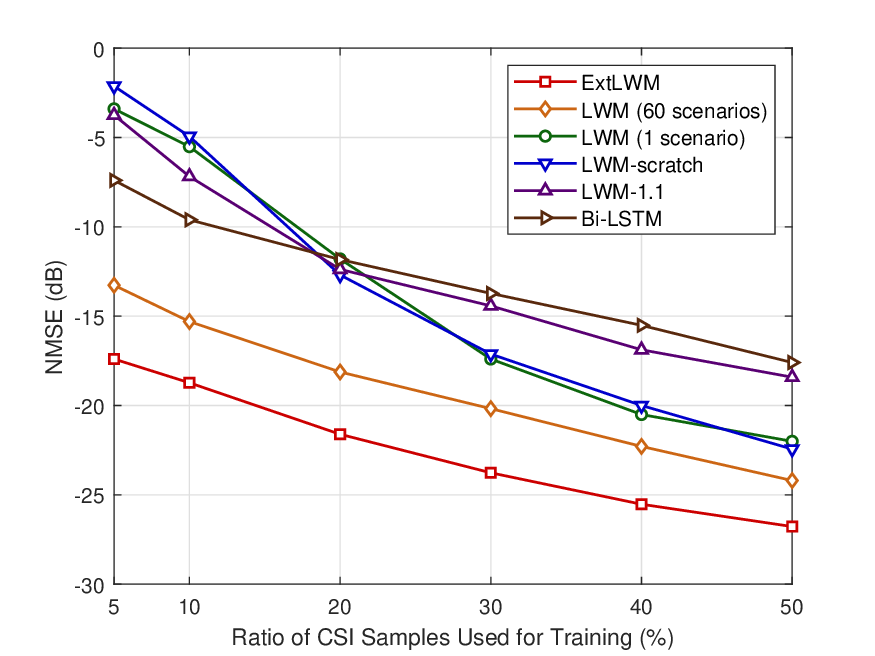}
\caption{Impact of training data volume.}
\label{fig_ext_sample}
\vspace{-0.3 cm}
\end{figure}

Fig. \ref{fig_ext_sample} illustrates the impact of data volume for training, indicated by training ratio $|\mathcal{K}_{\mathrm{ext}}|/1250$. Results show that ExtLWM exhibits increasingly pronounced performance gains over the model trained from scratch as the amount of training data decreases, achieving NMSE improvements of 8.9~dB, 9.1~dB, and 10~dB at training ratios of 20\%, 10\%, and 5\%, respectively.  Moreover, dedicated models typically capture only the propagation characteristics and spatial correlations of specific scenarios, making it difficult to effectively adapt to new channel distributions during cross-scenario transfer. When the amount of training data in the new scenario is further reduced, their performance also degrades more significantly. These results demonstrate that WFMs can achieve better CSI extrapolation performance in data-limited scenarios.

\vspace{-0.2cm}

\subsection{LoS/NLoS Identification Performance}\label{section:5-C}
In this subsection, we validate LoS/NLoS identification performance. The experiments are conducted using the data associated with ``BS3'' and 1250 users from row-500 in ``O1-blockage'' scenario from DeepMIMO-V3, with the system setting given by $M=180$, $\eta_{\mathrm{msk}}=90\%$, $f_{\mathrm{c}}=3.5$\,GHz, $S=16$, $D=\frac{\lambda_{\mathrm{c}}}{8}$. To simulate the sample-limited and label-scarce environment, 50\% of the users with complete CSI ($|\mathcal{K}_{\mathrm{ext}}|=625$) are used to fine-tune ExtLWM, while 1\% of users with known labels ($|\mathcal{K}_{\mathrm{id}}| = 12$) are used to train the connector. The LLM adopts ``GPT-2-small'' (124M parameters) from the GPT-2 family, with $D_{\mathrm{LLM}}=768$, $L_{\mathrm{dec}}=12$, and $H_{\mathrm{dec}}=12$. The connector is configured with $D_{\mathrm{q}}=3$ and $L_{\mathrm{q}}=2$ (0.4M parameters) \cite{OPENAI Radford}. The training hyperparameters are set as $E_{\mathrm{id}}=200$ and $\eta_{\mathrm{id}}=1\times10^{-3}$. The peak GPU memory consumption and training time for stage-3 are approximately $3.5$GB and $5$min, respectively.

To compare with the proposed multi-stage training strategy, we test other training schemes as follows:
\begin{itemize}
\item CLS embedding \cite{Arxiv Alikhani}: The LWM architecture supports the output of a single CLS embedding that efficiently aggregates information across the entire input. We replace the CSI embedding with the CLS embedding output by the fine-tuned ExtLWM for the identification task.

\item Joint fine-tuning \cite{ICCC Lu}: Based on the pre-trained ExtLWM, we jointly fine-tune all parameters, i.e., $\bm{\theta}_{\mathrm{WFM}}$ and $\bm{\theta}_{\mathrm{Con}}$, to optimize training loss of both extrapolation in \eqref{eq_loss_ext} and identification in \eqref{eq_loss_cls}, i.e., $\alpha_{\mathrm{ext}} L_{\mathrm{ext}} + \alpha_{\mathrm{id}} L_{\mathrm{id}}$.

\item Direct fine-tuning: We directly apply the pre-trained ExtLWM to the identification task, i.e., without first going through the extrapolation task to fine-tune $\bm{\theta}_{\mathrm{WFM}}$.

\item Raw CSI: We bypass ExtLWM and use only the raw incomplete CSI for the identification task.

\end{itemize}

\begin{figure}[t]
\centering
\subfigure[Lab.-CSI rat.: 1\%, $|\mathcal{K}_{\mathrm{id}}|=12$]{%
  \includegraphics[width=0.49\columnwidth]{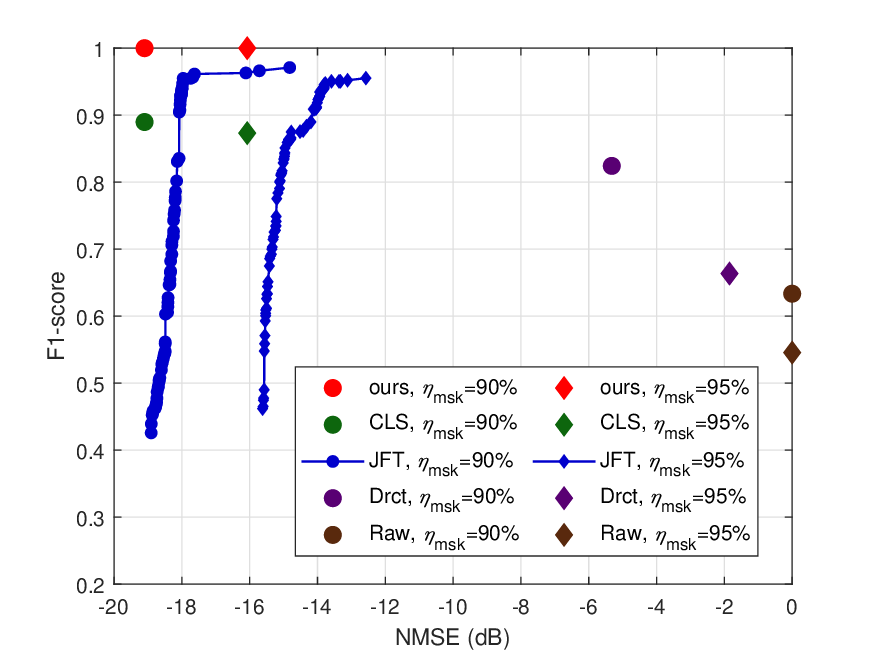}%
  \label{fig_cls_001}%
}
\hfill
\subfigure[Lab.-CSI rat.: 0.5\%, $|\mathcal{K}_{\mathrm{id}}|=6$]{%
  \includegraphics[width=0.49\columnwidth]{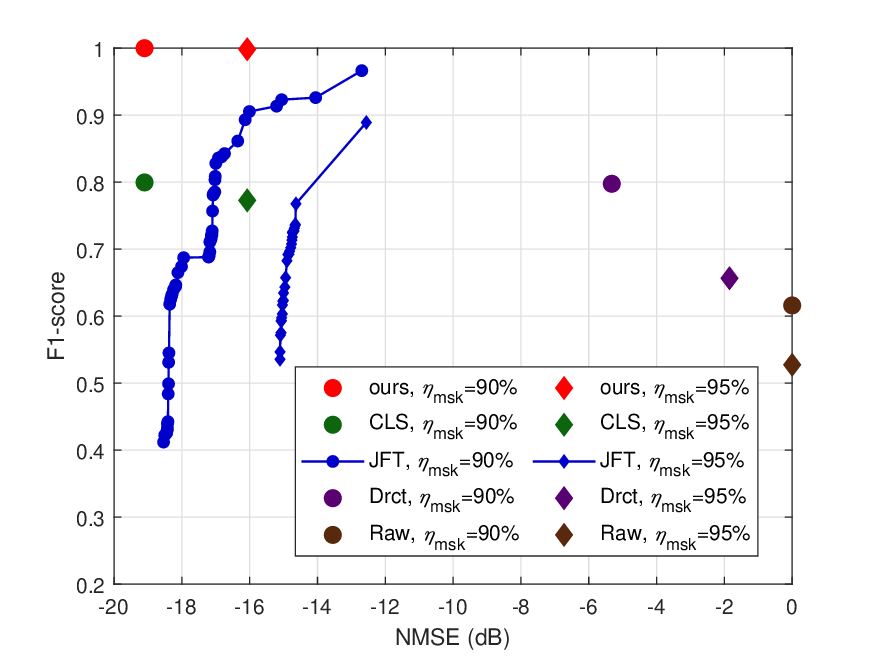}%
  \label{fig_cls_0005}%
}

\caption{Performance vs.\ labeled CSI ratio.}
\label{fig_cls_data}
\vspace{-0.2cm}
\end{figure}

In Fig.~\ref{fig_cls_data}, we plot both the F1-score of LoS/NLoS identification (on the y-axis) and the NMSE of CSI extrapolation (on the x-axis) obtained by the aforementioned training schemes using different ratios of LoS/NLoS-labeled training samples. Several key observations are elaborated in the sequel. 1)~Firstly, even with very limited labeled CSI, i.e., $|\mathcal{K}_{\mathrm{id}}| = 6$, our multi-stage training delivers F1-score of 1.00 and 0.998, respectively, at high CSI incompleteness ratios of $\eta_{\mathrm{msk}}=90\%$ and $\eta_{\mathrm{msk}}=95\%$, underscoring its superior robustness in extremely label-scarce regimes. 2)~Secondly, as shown in Fig. \ref{fig_cls_001}, replacing CSI embeddings with CLS leads to F1-score degradations of 12.7\% and 11\%. This gap widens to 22.7\% and 22.0\% for fewer labels in Fig. \ref{fig_cls_0005}. This is due to the fact that incomplete CSI renders CLS representations susceptible to capturing missing/noisy patterns over salient physical features, impairing its classification capability. 3)~Finally, the Pareto frontier of joint fine-tuning is depicted in Fig. \ref{fig_cls_data}, which is dominated by our proposed multi-stage training paradigm. This sub-optimality arises because acute label scarcity intensifies inter-task gradient conflict and exacerbates task-dominance imbalances.

\begin{figure}[t]
\centering

\subfigure[Raw CSI]{%
  \includegraphics[width=0.33\columnwidth]{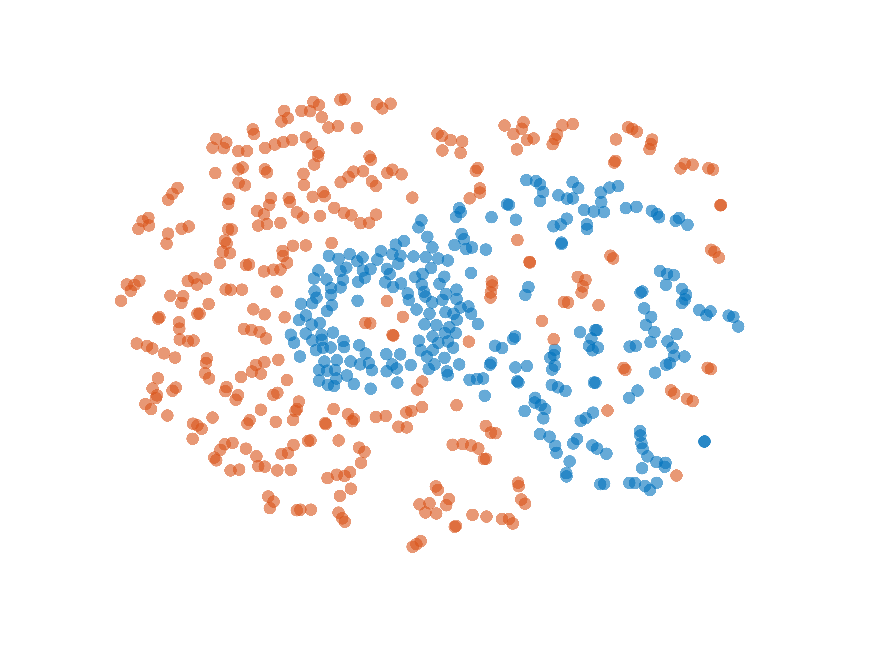}%
  \label{fig_tsne_raw}%
}
\hspace{-0.3cm}
\subfigure[w/ CSI extrap.]{%
  \includegraphics[width=0.33\columnwidth]{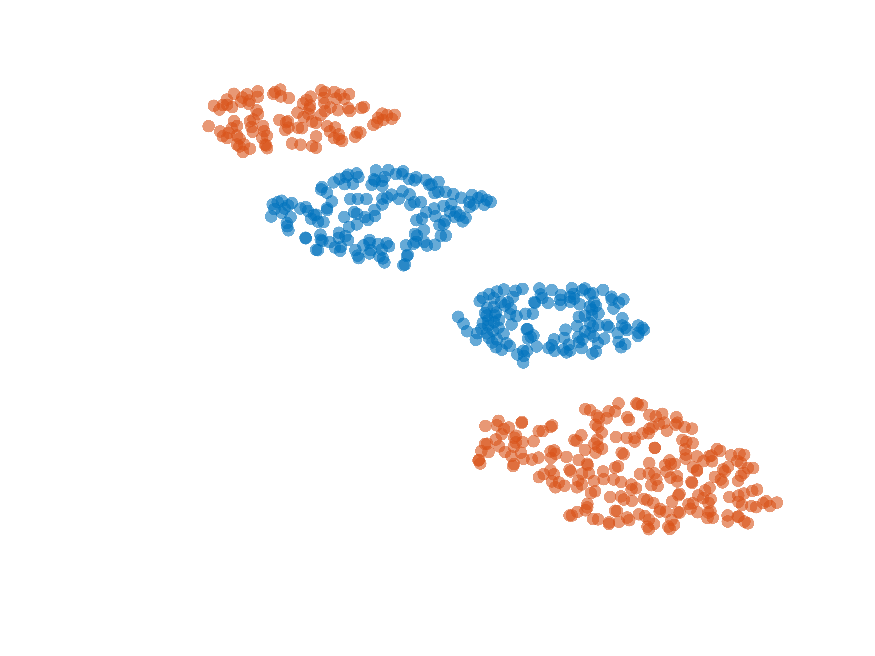}%
  \label{fig_tsne_csi_emb}%
}
\hspace{-0.3cm}
\subfigure[w/o CSI extrap.]{%
  \includegraphics[width=0.33\columnwidth]{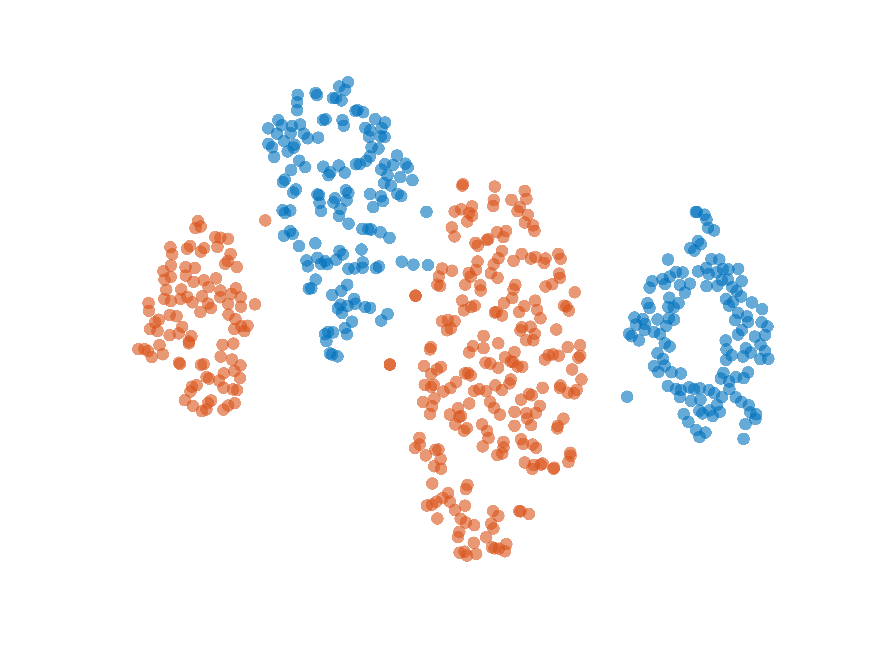}%
  \label{fig_tsne_org}%
}

\caption{Impact of CSI extrapolation on embedding.}
\label{fig_cls_tsne}
\vspace{-0.3cm}
\end{figure}

Fig. \ref{fig_cls_tsne} provides $t$-SNE projections of CSI embeddings comparing the post-extrapolated ExtLWM (Fig. \ref{fig_tsne_csi_emb}) with the pre-trained counterpart (Fig. \ref{fig_tsne_org}) to visualize the impact of extrapolation task on LoS/NLoS identification. The results reveal that integrating CSI extrapolation stimulates the model to distill latent features of MA channels more effectively and to yield more discriminative CSI representations that substantially bolster the performance of downstream tasks.

\vspace{-0.2cm}

\subsection{Sum-Rate Maximization Performance}
In this subsection, we proceed to examine the effectiveness of RL training in resolving the sum-rate maximization task. In addition to the system settings specified in Sec.~\ref{section:5-C}, we set $d_\mathrm{min}=\frac{\lambda_{\mathrm{c}}}{2}$, $N=8$, $P_{\mathrm{tx}}=20$dBm, $\sigma_{\mathrm{c}}^2=-70$dBm, and $K=6$. For RL algorithm, we set epoch $E_{\mathrm{max}}=500$, learning rate $\eta_{\mathrm{max}}=3\times10^{-4}$, $\epsilon_{\mathrm{c}}=0.2$, $G=20$, and $|\mathcal{B}_{\mathrm{sub}}|= \frac{GN}{4}$, while $\lambda_{\mathrm{exp}}$ is annealed from 0.1 to 0.02. To implement ER, we set $\lambda_{\mathrm{cl}}=5\times10^{-3}$ and $\mathcal{K}_{\mathrm{cl}}=\mathcal{K}_{\mathrm{id}}$. The average execution times of the proposed framework for CSI extrapolation, LoS/NLoS identification, and sum-rate maximization are approximately $0.016$s, $0.025$s, and $0.052$s, respectively. The peak GPU memory consumption and training time for stage-4 are approximately $10.3$GB and $2.3$h, respectively.

\begin{figure}[t]
\centering
\includegraphics[width = 2.8 in]{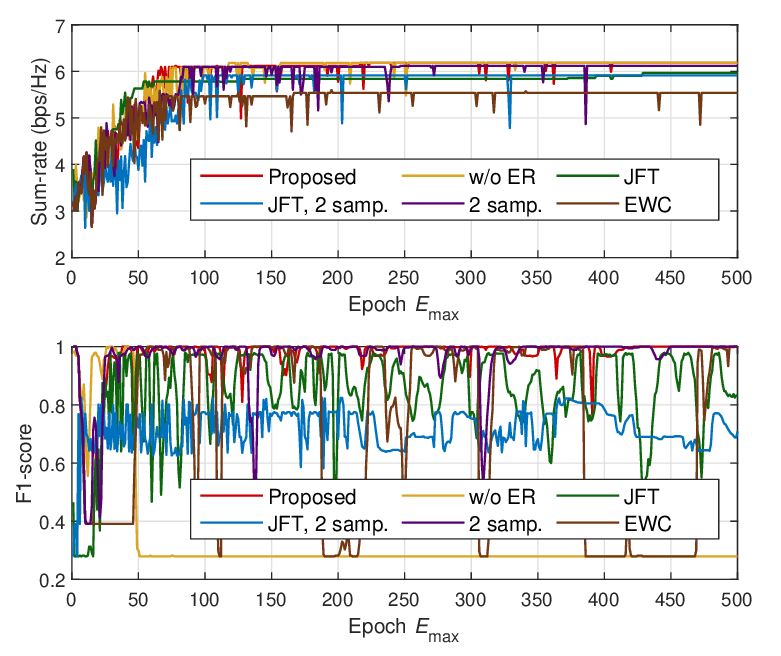}
\caption{Performance of ER method.}
\label{fig_max_cl}
\vspace{-0.3 cm}
\end{figure}

In Fig. \ref{fig_max_cl}, we evaluate the efficacy of ER in maintaining LoS/NLoS identification precision using the validation set in Sec. \ref{section:5-C}. Results show that ER sustains an F1-score of 1.00 during sum-rate optimization. In contrast, omitting ER leads to catastrophic forgetting at epoch 49, with model losing its previously held identification accuracy, which never recovers. Notably, even with $|\mathcal{K}_{\mathrm{cl}}|$ reduced from 12 to 2 (``2 samp'' in legend), ER still preserves identification capabilities perfectly, whereas skipping the preliminary fine-tuning on the identification task (denoted as ``JFT, 2 samp.'' in the legend) results in an F1-score of only 0.824.
These results demonstrate the effectiveness of the ER-based strategy and its strong training-sample efficiency.
Moreover, we compare the joint fine-tuning scheme (denoted as ``JFT''), which skips stage-3 and directly optimizes LoS/NLoS identification during sum-rate maximization. The results show that its sum-rate performance degrades noticeably, while the maximum F1-score reaches only 0.979. Together with the results in Fig.~8, these experiments further demonstrate the effectiveness of the proposed multi-stage training strategy.
Finally, we compare ER with elastic weight consolidation (EWC) algorithm~\cite{PNAS Kirkpatrick}, which mitigates forgetting by constraining parameters important to previous tasks. Experimental results show that such constraints inhibit sum-rate gains, validating the superior performance trade-off achieved by ER.

\begin{figure}[t]
\centering
\includegraphics[width = 2.8 in]{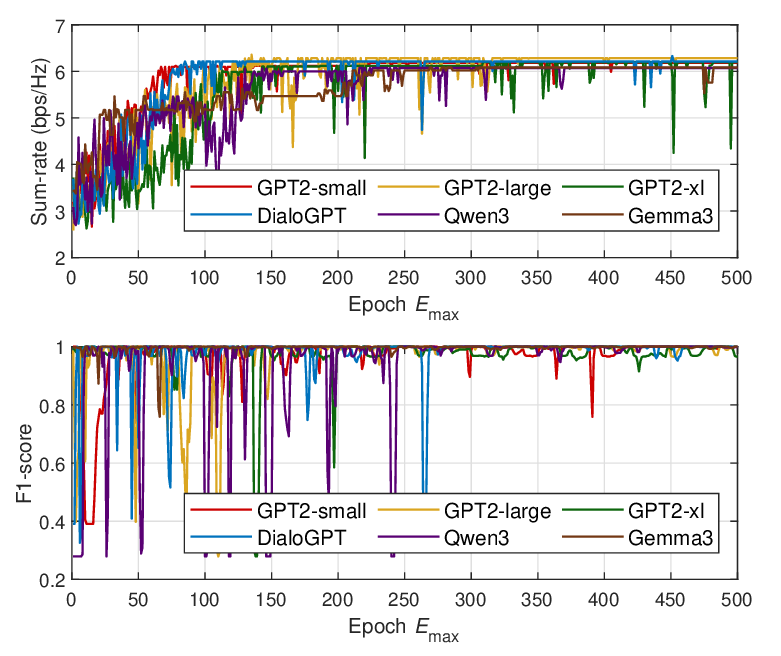}
\caption{Performance of diverse LLMs.}
\label{fig_max_LLM}
\vspace{-0.3 cm}
\end{figure}

Fig.~\ref{fig_max_LLM} validates the effectiveness and generalizability of the proposed multi-stage training paradigm. We first deploy the proposed framework on GPT-2-large (774M) and GPT-2-xl (1.5B), which adopt different configurations of $D_{\mathrm{LLM}}$, $H_{\mathrm{dec}}$, and $L_{\mathrm{dec}}$. The results demonstrate that the proposed training strategy can achieve satisfactory performance across LLMs with similar architectures but different model scales. We then further consider DialoGPT (124M)~\cite{ACL Zhang}, a dialogue-oriented variant fine-tuned from GPT-2, and show that the proposed training framework can also be directly transferred to task-specialized LLMs. Finally, we further consider Qwen3 (0.6B)~\cite{Arxiv Yang} and Gemma3 (270M)~\cite{Arxiv Gemma3}, demonstrating that the proposed modality alignment mechanism \eqref{eq_multi_emb}, modified causal mask \eqref{eq_mha_lwm}, and corresponding training strategy remain applicable to LLMs with different architectures\footnote{Notably, the tokenizers of Qwen3 and Gemma3 support only the digits $0$--$9$ as single-token numerical representations, and thus cannot directly cover all candidate positions in the considered MA system. To address this issue, we extend $\mathcal{T}_{\max}$ to the set of token IDs corresponding to $\{\text{``}0\text{''},\ldots,\text{``}9\text{''},[\mathrm{SPACE}]\}$, where $[\mathrm{SPACE}]$ acts as a termination token for the current position, enabling each candidate position to be represented using at most three tokens. Accordingly, while preserving the feasible position set determined by the original action mask \eqref{eq_action_mask}, i.e., $\{p_n\mid M_{\min,n}\leq p_n \leq M_{\max,n}\}$, we further introduce a prefix-dependent token-level action mask. Specifically, at each decoding step, only the digit tokens that can extend the current prefix to at least one feasible position are allowed, while $[\mathrm{SPACE}]$ is allowed only when the current prefix itself forms a feasible position. The results demonstrate that the proposed multi-token encoding scheme can effectively accommodate different LLM tokenizers and may further alleviate the limitation of GPT-2's single-token numerical range in large-scale candidate-position spaces.}.

\begin{figure}[t]
\centering
\includegraphics[width = 3.0 in]{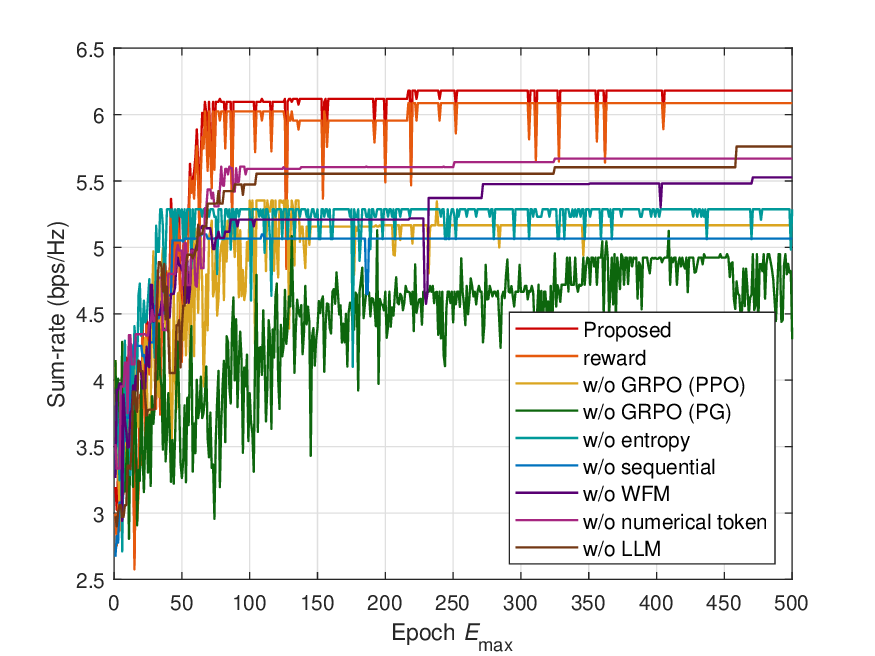}
\caption{Performance of RL method.}
\label{fig_curves_benchmarks}
\vspace{-0.3 cm}
\end{figure}

\begin{table}[t]
    \centering
    \footnotesize
    \caption{Overall comparison of different precoders.}
    \label{table_precoder}
    \begin{tabular}{|c|c|c|c|}
    \hline
    Precoder & WMMSE & MMSE & ZF \\
    \hline
    Execution time (s) & 0.487 & 0.052 & 0.052 \\
    \hline
    Sum-rate (bps/Hz) & 9.684 & 6.180 & 1.443 \\
    \hline
    \end{tabular}
    \vspace{-0.2cm}
\end{table}

In Fig.~\ref{fig_curves_benchmarks}, we validate the effectiveness of the proposed RL method from the following aspects: 1)~The designed reward function can effectively guide the improvement of the actual sum-rate, further validating that ExtLWM can generate high-quality extrapolated CSI and feature embeddings in a self-contained manner to effectively guide MA positioning.~2)~Removing the entropy loss term in \eqref{eq_loss_exp} results in a 14.45\% sum-rate degradation, demonstrating the importance of balancing exploration and exploitation during training. 3)~Compared with PPO~\cite{Arxiv Schulman} and the classical policy gradient (PG) method, i.e., REINFORCE~\cite{ML Williams}, GRPO achieves clearly better performance, mainly benefiting from its group-relative advantage estimation, which can effectively alleviate the credit-assignment difficulty. Meanwhile, replacing numerical tokens with random tokens leads to an 8.28\% sum-rate degradation. These results demonstrate that the proposed design can better exploit the potential of LLMs for sequential MA positioning. 4)~Replacing the frozen ``GPT-2-small'' with an 8.45M-parameter fully trained transformer instead results in a 6.8\% sum-rate degradation. In contrast, the proposed method only trains the 0.4M parameters of the connector, demonstrating its ability to effectively exploit the trainable-parameter efficiency of LLMs. 5)~Removing the pre-trained weights of ExtLWM and training it from scratch only on the downstream scenario leads to a 10.68\% sum-rate degradation, a 3.84~dB degradation in CSI extrapolation performance, and a reduction of the F1-score to 96.6\%. These results demonstrate that integrating the WFM with the LLM can better support data-limited MA scenarios. 6)~Replacing the WTE of LLM with $N$ $M$-class embeddings transforms MA positioning into a single-shot parallel prediction based solely on $\mathbf{Z}_{\mathrm{csi}}$, resulting in an 18.03\% degradation in sum-rate performance. This demonstrates that sequential generation increases the serial computational depth of the LLM, thereby enabling it to better handle problems with sequential dependencies under a fixed network depth.

In Table \ref{table_precoder}, we compare the sum-rate performance and execution time achieved under different precoding schemes. The results show that different precoders can effectively serve as part of the reward function to guide the RL agent. Specifically, WMMSE achieves an approximately 56.7\% sum-rate gain over MMSE, but requires approximately 9.37 times the execution time of MMSE. This substantially higher computational overhead mainly arises from the multiple alternating iterations over auxiliary variables required by WMMSE, together with the one-dimensional search performed in each beamforming update to determine the Lagrange multiplier satisfying the transmit power constraint. In contrast, MMSE achieves an approximately 328.3\% sum-rate improvement over ZF, demonstrating its effectiveness in suppressing multiuser interference under correlated channel conditions. Therefore, considering the overall trade-off between achievable sum-rate and computational complexity, we choose MMSE precoder in the sum-rate maximization task.

Next, we compare several MA positioning algorithms for sum-rate maximization, as summarized below. For fairness, all benchmark schemes adopt the MMSE precoder \eqref{eq_MMSE}.

\begin{itemize}
\item BnB~\cite{TCOM Wu} performs a systematic tree search over the discrete antenna position space.

\item PBR~\cite{CL Cang} relaxes binary variables to continuous ones and enforces discreteness through penalty terms.

\item Greedy sequentially selects the optimal MA positions based on locally maximum sum-rate.

\item FPA assumes a fixed antenna spacing of $\frac{\lambda_{\mathrm{c}}}{2}$.

\end{itemize}

\begin{figure}[t]
\centering
\includegraphics[width = 3.0 in]{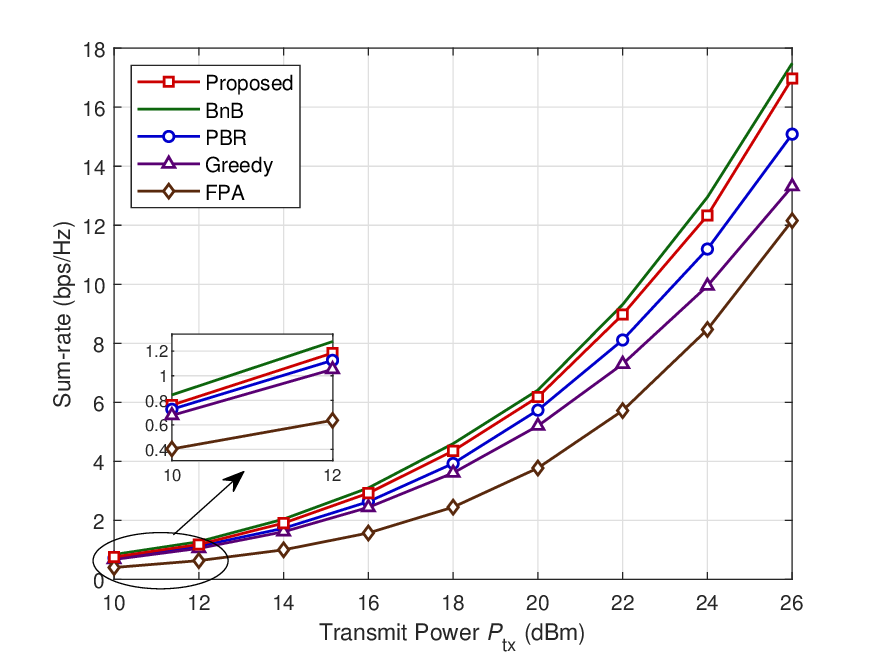}
\caption{Impact of $P_{\mathrm{tx}}$ on sum-rate.}
\label{fig_max_P}
\vspace{-0.3 cm}
\end{figure}

Fig.~\ref{fig_max_P} illustrates the sum-rate performance achieved by different methods. The results show that sequential optimization can achieve satisfactory performance for MA positioning. Specifically, at $P_{\mathrm{tx}}=10$ dBm and $P_{\mathrm{tx}}=26$ dBm, the proposed scheme achieves $90.17\%$ and $97.05\%$ of the sum rate obtained by the BnB method, respectively. This indicates that sequential optimization indeed incurs a certain degree of suboptimality compared with directly searching the original combinatorial space. On the other hand, it outperforms the PBR method, mainly because the relaxation and penalty-based approximation may introduce a performance gap from the original combinatorial optimization problem.

\begin{figure}[t]
\centering
\includegraphics[width = 3.0 in]{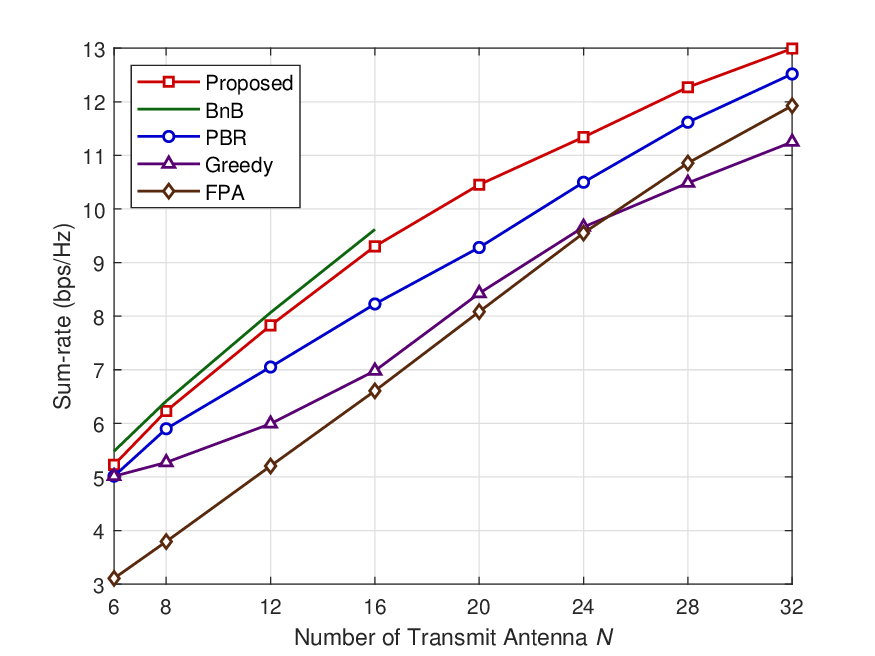}
\caption{Impact of $N$ on sum-rate.}
\label{fig_max_N}
\vspace{-0.3 cm}
\end{figure}

Next, we compare overall performance of different algorithms under different $N$, with the results presented in Fig.~\ref{fig_max_N} and Table~\ref{table_max_N}, respectively. The results show that, by autoregressively generating MA positions based on the learned semantic representations, the proposed method achieves approximately $96.6\%$ of the upper-bound sum-rate performance. In contrast, the search space of BnB grows combinatorially with $N$, leading to a rapidly increasing execution time and limiting its applicability to large-scale MA configurations. Compared with PBR, the proposed method achieves an average sum-rate gain of approximately $8.0\%$, while avoiding the substantial computational overhead caused by repeated alternating optimization. Although Greedy achieves the lowest execution time, its myopic position selection strategy may lead to accumulated suboptimal decisions, resulting in noticeable performance degradation as $N$ increases. Overall, these results demonstrate that the proposed framework can achieve a favorable trade-off between sum-rate performance and computational efficiency.

\begin{table}[t]
    \centering
    \footnotesize
    \caption{Impact of $N$ on execution time.}
    \label{table_max_N}
    \begin{tabular}{|c|c|c|c|c|}
    \hline
    $N$ & Proposed & BnB & PBR & Greedy \\
    \hline
    6  & 0.043s & 553.659s & 35.349s  & 0.016s \\
    \hline
    8  & 0.052s & 0.443h   & 41.553s  & 0.017s \\
    \hline
    12 & 0.066s & 4.193h   & 55.382s  & 0.018s \\
    \hline
    16 & 0.083s & 58.031h  & 71.264s  & 0.019s \\
    \hline
    20 & 0.095s & --       & 82.720s  & 0.021s \\
    \hline
    24 & 0.113s & --       & 93.366s  & 0.024s \\
    \hline
    28 & 0.129s & --       & 106.527s & 0.025s \\
    \hline
    32 & 0.144s & --       & 111.746s & 0.032s \\
    \hline
    \end{tabular}
    \vspace{-0.2cm}
\end{table}

\begin{table}[t]
    \centering
    \footnotesize
    \caption{Impact of $M$ on execution time and sum-rate.}
    \label{table_max_M}
    \begin{tabular}{|c|c|c|c|c|c|c|}
    \hline
    \multirow{2}{*}{$M$}
    & \multicolumn{3}{c|}{Execution time (s)}
    & \multicolumn{3}{c|}{Sum-rate (bps/Hz)} \\
    \cline{2-7}
    & Proposed & PBR & Greedy
    & Proposed & PBR & Greedy \\
    \hline
    100 & 0.089 & 47.901 & 0.018 & 8.885 & 8.662 & 8.120 \\
    \hline
    140 & 0.092 & 67.840 & 0.019 & 9.631 & 8.977 & 8.300 \\
    \hline
    180 & 0.095 & 82.720 & 0.021 & 10.451 & 9.281 & 8.426 \\
    \hline
    220 & 0.099 & 106.807 & 0.028 & 10.924 & 9.420 &8.572 \\
    \hline
    260 & 0.104 & 139.765 & 0.030 & 11.138 & 9.609 & 8.641\\
    \hline
    \end{tabular}
    \vspace{-0.2cm}
\end{table}

Finally, Table~\ref{table_max_M} presents the overall performance of different algorithms under different $M$, where $\eta_{\text{msk}} = 80\%$ and $N = 20$. The results show that, with scenario-specific fine-tuning of ExtLWM, the proposed framework can effectively learn MA channel characteristics under different candidate-position spaces. As $M$ increases, the growth in overall inference latency becomes noticeably milder than that observed with increasing $N$, because ExtLWM, rather than GPT-2, becomes the dominant contributor to the computational overhead. In addition, the proposed multi-stage training algorithm maintains a stable average CSI extrapolation NMSE of approximately $-24.6$dB, while ER consistently preserves the LoS/NLoS identification performance with an F1-score of $1.0$. These results further demonstrate the robustness and scalability of the proposed algorithm.

\vspace{-0.2cm}

\section{Conclusions}\label{section:6}

This paper proposes a unified ExtLWM-Connector-GPT architecture for multi-task processing in MA systems. Simulation results show that ExtLWM outperforms dedicated models and models trained from scratch in downstream CSI extrapolation. Meanwhile, the proposed approach effectively activates the sequential optimization capability of the LLM and achieves a favorable performance--complexity trade-off compared with conventional alternatives.
In future work, we will investigate the integration of the proposed framework into multimodal semantic communication systems to extend its applicability. Moreover, we will explore partially unfreezing the LLM while preserving existing task performance to further investigate its potential performance gains.

\bibliographystyle{IEEEtran}
\begin{IEEEbiography}[{\includegraphics[width=1.25in,height=1.25in,clip,keepaspectratio]{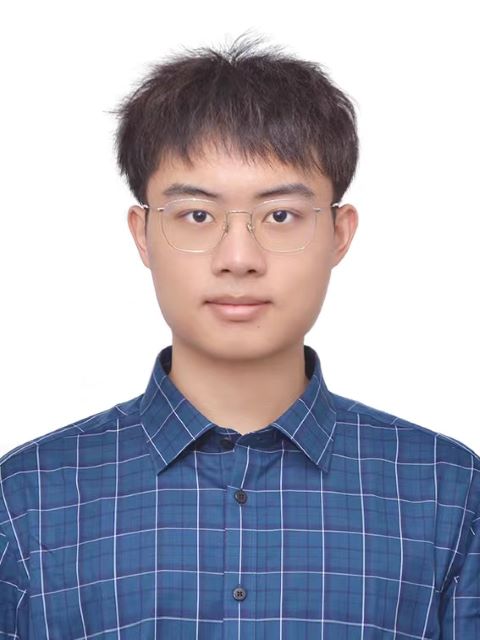}}]
{Haocheng Zhang} received the B.S. degree in electronic information engineering from Dalian University of Technology, Dalian, China, in 2022. He is currently pursuing the Ph.D. degree with the School of Information and Communication Engineering, Dalian University of Technology, Dalian, China. His current research interests focus on beamforming optimization based on deep reinforcement learning.
\end{IEEEbiography}

\bibliographystyle{IEEEtran}
\begin{IEEEbiography}[{\includegraphics[width=1.25in,height=1.25in,clip,keepaspectratio]{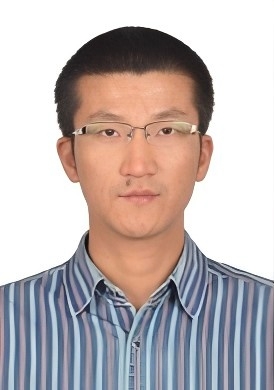}}]
{Yang Liu} (Member, IEEE) received the Ph.D. degree in Electrical Engineering from Lehigh University, PA, USA, in 2016. Before that he received the B.E. and M.E. degrees in Electrical Engineering from Beijing University of Posts and Telecommunications (BUPT), Beijing, China, in 2007 and 2010, respectively. In 2016, he worked as a senior engineer in Marvell Semiconductor Ltd., Santa Clara, CA, USA. From 2016 to 2019, he worked as a research engineer in Huawei Technologies Co., Ltd., Shanghai, China. From 2019 to 2020, he worked as a postdoctoral research fellow in Nanyang Technological University (NTU), Singapore. Since April 2020, he has been with School of Information and Communication Engineering, Dalian University of Technology (DUT), Dalian, China. His research interests include MIMO technology, RIS, ISAC, decentralized signal processing and signal detection and estimation.
\end{IEEEbiography}

\bibliographystyle{IEEEtran}
\begin{IEEEbiography}[{\includegraphics[width=1.25in,height=1.25in,clip,keepaspectratio]{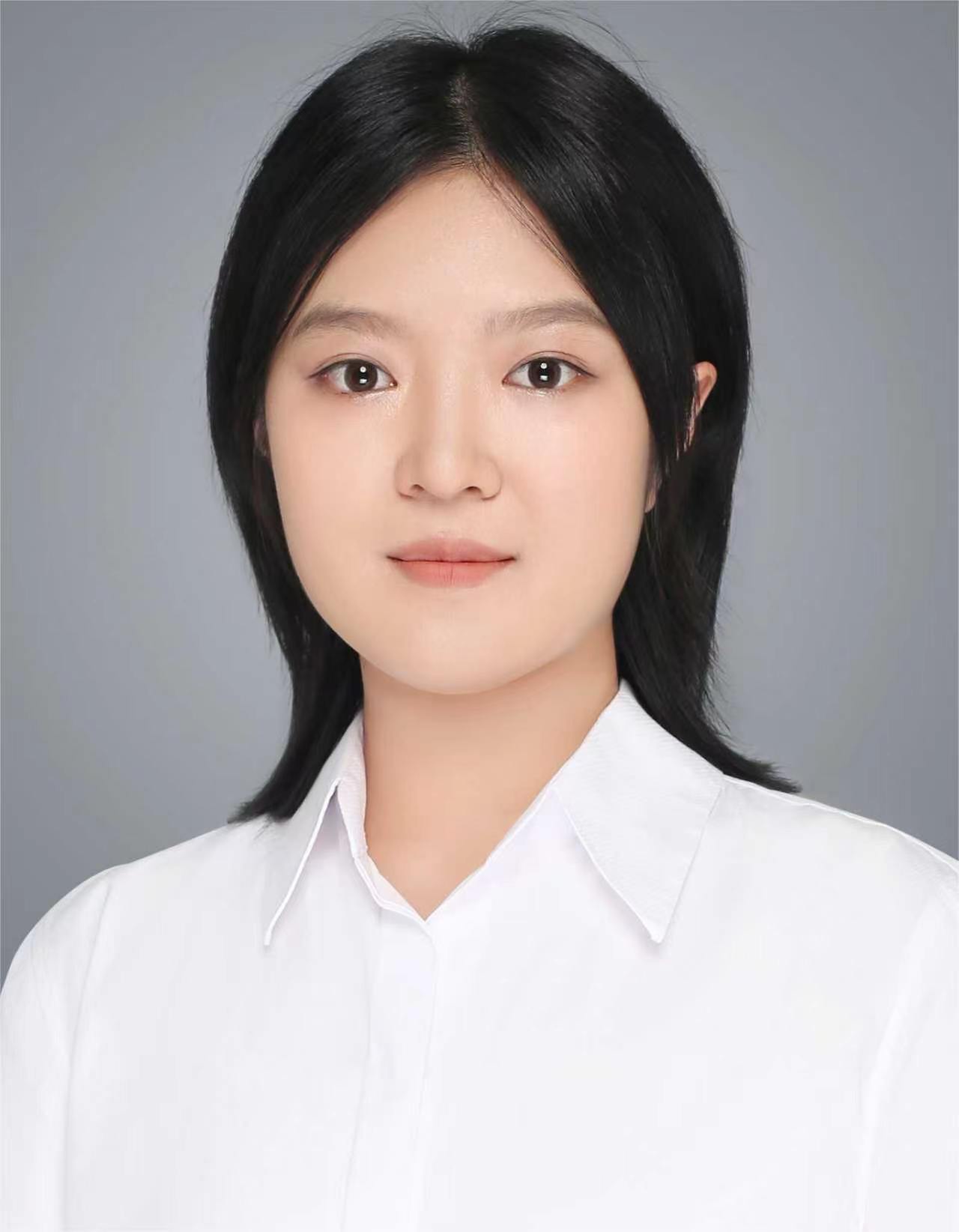}}]
{Mengzhen Liu} received the B.S. degree in electronic information engineering from Dalian University of Technology, Dalian, China, in 2023. She is now studying toward the Ph.D. degree with the School of Information and Communication Engineering, Dalian University of Technology, Dalian, China. 
Her current research interests are focused on reconfigurable antennas, extremely large-scale antenna array systems, and mmWave communications.
\end{IEEEbiography}

\bibliographystyle{IEEEtran}
\begin{IEEEbiography}[{\includegraphics[width=1.25in,height=1.25in,clip,keepaspectratio]{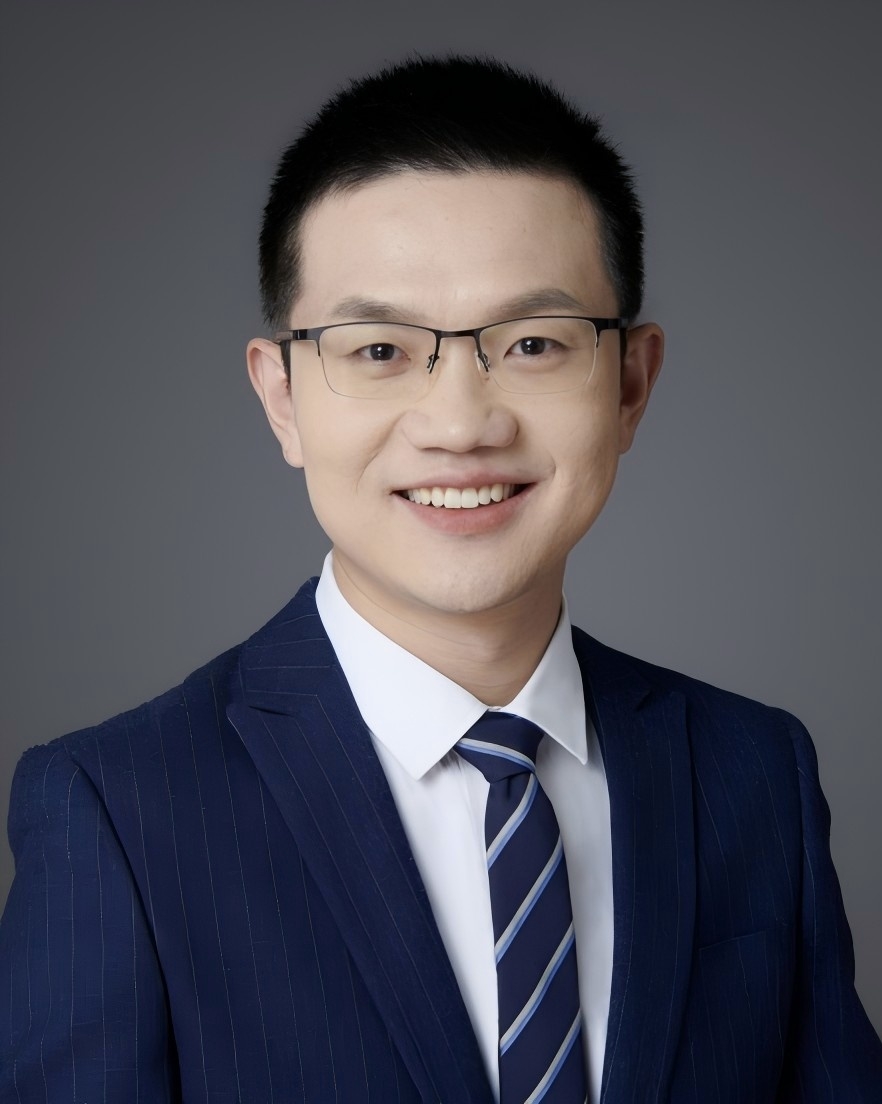}}]
{Qingqing Wu} (Senior Member, IEEE) is currently a Full Professor with the School of Information Science and Electronic Engineering, Shanghai Jiao Tong University, China. His current research interest includes intelligent reflecting surface (IRS), unmanned aerial vehicle (UAV) communications, and MIMO transceiver design. He has coauthored more than 200 IEEE journal papers with 40+ ESI highly cited papers, which have received more than 60,000 Google citations. He has been listed as the Clarivate ESI Highly Cited Researcher since 2021.

He serves as an Associate/Senior/Area Editor for IEEE Transactions on Wireless Communications, IEEE Transactions on Communications, IEEE Communications Letters, IEEE Wireless Communications Letters. He is the Lead Guest Editor for IEEE Journal on Selected Areas in Communications. He is the workshop co-chair for IEEE ICC 2019-2023 and IEEE GLOBECOM 2020. He is the Founding Chair of IEEE ComSoc Asia-Pacific Young Professional committee and Chair of IEEE VTS Drone Committee.
\end{IEEEbiography}

\bibliographystyle{IEEEtran}
\begin{IEEEbiography}[{\includegraphics[width=1.25in,height=1.25in,clip,keepaspectratio]{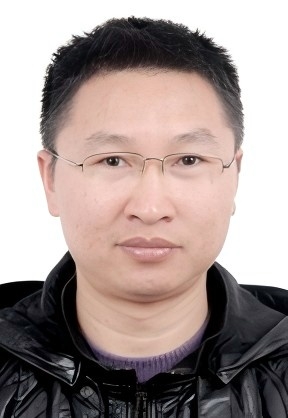}}]
{Qingjiang Shi} (Member, IEEE) received the Ph.D. degree in electronic engineering from Shanghai Jiao Tong University, Shanghai, China, in 2011. From September 2009 to September 2010, he visited Prof. Z.-Q. (Tom) Luo's research group at the University of Minnesota, Twin Cities. In 2011, he worked as a Research Scientist at Bell Labs China. From 2012, he was with the School of Information and Science Technology at Zhejiang Sci-Tech University. From Feb. 2016 to Mar. 2017, he worked as a research fellow at Iowa State University, USA. From 2018, he is currently a professor with the School of Computer Science and Technology at Tongji University. His interests lie in algorithm design and analysis with applications in machine learning, signal processing and wireless networks. So far he has published more than 100+ IEEE journals and filed more than 50 national patents.

Dr. Shi has served on the editorial board of IEEE TRANSACTIONS ON SIGNAL PROCESSING as an Associate Editor. He was the recipient of the IEEE Signal Processing Society Best Paper Award in 2022, the Second Prize of Zhejiang Provincial Natural Science Award in 2023, the Excellent Technical Cooperation Award From Huawei Wireless Network Product Line in 2024, the Huawei Technical Cooperation Achievement Transformation Award (2nd Prize) in 2022, the Huawei Outstanding Technical Achievement Award in 2021, the Golden Medal at the 46th International Exhibition of Inventions of Geneva in 2018, the First Prize of Science and Technology Award from China Institute of Communications in 2017, the National Excellent Doctoral Dissertation Nomination Award in 2013, the Shanghai Excellent Doctoral Dissertation Award in 2012, and the Best Paper Award from the IEEE PIMRC'09 conference.
\end{IEEEbiography}

\bibliographystyle{IEEEtran}
\begin{IEEEbiography}[{\includegraphics[width=1.25in,height=1.25in,clip,keepaspectratio]{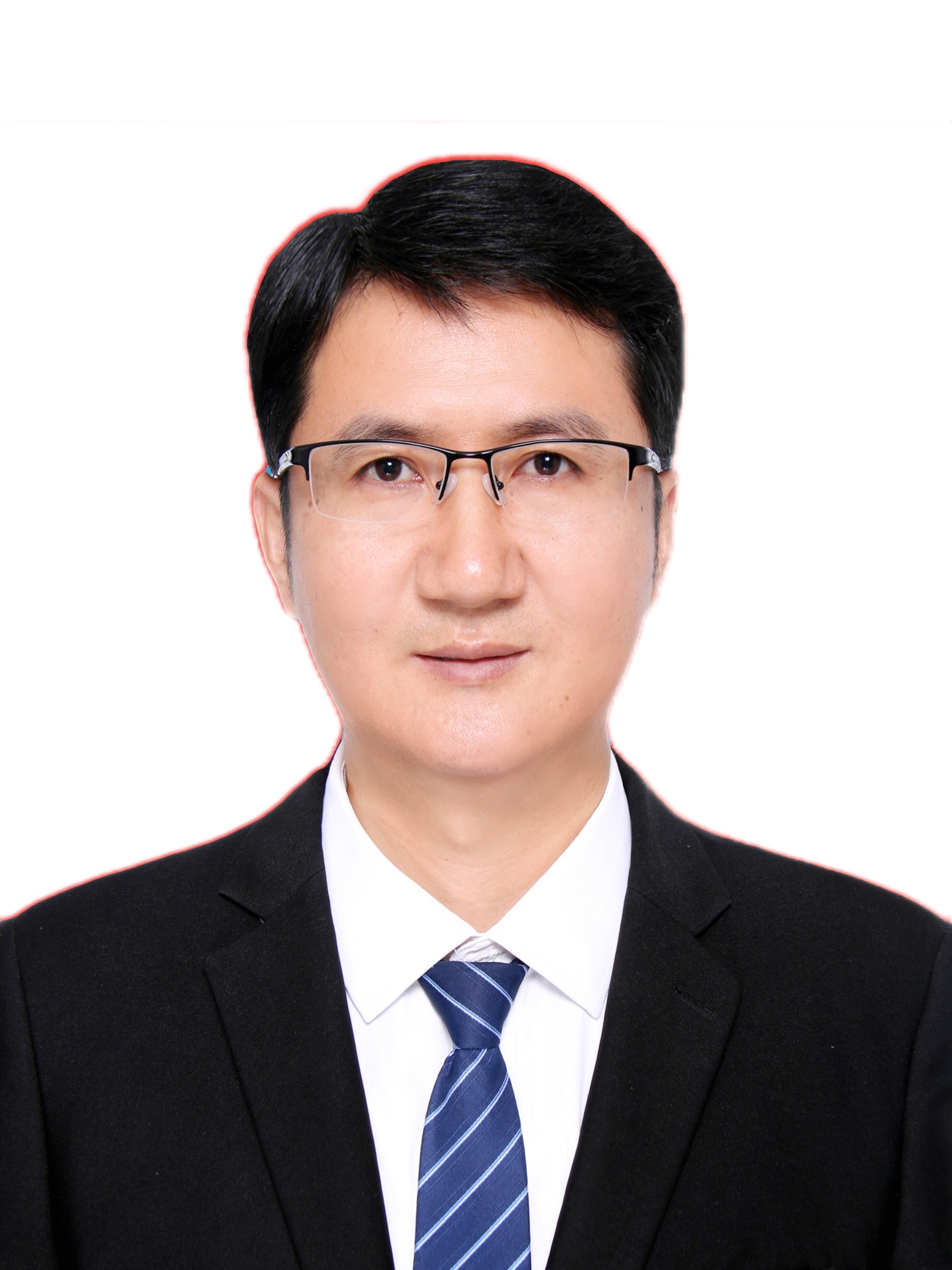}}]
{Wei Wang} (Senior Member, IEEE) received the Ph.D. degree in navigation, guidance, and control from Harbin Engineering University (HEU), Heilongjiang, China, in 2005. He was a Post-Doctoral Research Associate with the Harbin Institute of Technology, China, from July 2006 to April 2009. He was also an Associate Professor with HEU from August 2008 to August 2010 and an Academic Visitor with Loughborough University, U.K., from January 2010 to December 2010. He has been a Professor with HEU since September 2011. He has published about more than 80 refereed journals and conference papers. His current research interests include signal processing for wireless navigation systems and MIMO radar.
\end{IEEEbiography}

\end{document}